%% file: main.tex
\documentclass[aps,prd,superscriptaddress,floatfix,nofootinbib]{revtex4}

\usepackage{verbatim}
\usepackage{mathtools}
\usepackage{ragged2e}
\usepackage{amsmath}
\usepackage{amsfonts}
\usepackage{amssymb}
\usepackage{graphicx}
\usepackage{slashed}
\usepackage{bm}
\usepackage{color}
\usepackage{epsf}
\usepackage{orcidlink}
\usepackage[parfill]{parskip}
\usepackage{nicefrac}
\usepackage{mathrsfs}

\usepackage{hyperref}
\hypersetup{
  colorlinks   = true, 
  urlcolor     = blue, 
  linkcolor    = black, 
  citecolor   = red 
}

\begin{document}

\title{Timelike Compton scattering on a spin-0 target with kinematic twist-4~precision}

\author{V.~Mart\'inez-Fern\'andez\,\orcidlink{0000-0002-0581-7154}}
\affiliation{National Centre for Nuclear Research (NCBJ), 02-093 Warsaw, Poland}
\affiliation{IRFU, CEA, Universit\'e Paris-Saclay, F-91191 Gif-sur-Yvette, France}
\affiliation{Center for Frontiers in Nuclear Science, Stony Brook University, Stony Brook, NY 11794, USA}
\author{B.~Pire\,\orcidlink{0000-0003-4882-7800}}
\affiliation{Centre de Physique Th\'eorique, CNRS, École Polytechnique, I.P. Paris, 91128 Palaiseau, France  }

\author{P.~Sznajder\,\orcidlink{0000-0002-2684-803X}}
\affiliation{National Centre for Nuclear Research (NCBJ), 02-093 Warsaw, Poland}

\author{J.~Wagner\,\orcidlink{0000-0001-8335-7096}}
\affiliation{National Centre for Nuclear Research (NCBJ), 02-093 Warsaw, Poland}

\date{\today}

\newcommand{\complex}{\mathbb{C}}
\newcommand{\R}{\mathbb{R}}
\newcommand{\N}{\mathbb{N}}
\newcommand{\Z}{\mathbb{Z}}
\newcommand{\T}{\mathcal{T}}
\newcommand{\quark}{\mathfrak{q}}
\newcommand{\lag}{\mathscr{L}}
\newcommand{\smatrix}{\mathcal{S}}
\newcommand{\g}{{\rm g}_{\rm s}}
\newcommand{\D}{\mathscr{D}}
\newcommand{\W}{\mathcal{W}}
\newcommand{\re}{\mathfrak{Re}}
\newcommand{\im}{\mathfrak{Im}}
\newcommand{\eps}{\varepsilon} 
\newcommand{\M}{\mathcal{M}} 
\newcommand{\amp}{\mathcal{A}} 
\newcommand{\rt}{\mathscr{R}}
\newcommand{\Ii}{I_{\rm (i)}}
\newcommand{\Iii}{I_{\rm (ii)}}
\newcommand{\Iiii}{I_{\rm (iii)}}
\newcommand{\Iiv}{I_{\rm (iv)}}
\newcommand{\Iv}{I_{\rm (v)}}
\newcommand{\Ivi}{I_{\rm (vi)}}
\newcommand{\Ivii}{I_{\rm (vii)}}
\newcommand{\Iviii}{I_{\rm (viii)}}
\newcommand{\pbbi}{\mathbb{P}_{\rm (i)}}
\newcommand{\pbbii}{\mathbb{P}_{\rm (ii)}}
\newcommand{\pbbiii}{\mathbb{P}_{\rm (iii)}}
\newcommand{\pbbiv}{\mathbb{P}_{\rm (iv)}}
\newcommand{\pbbv}{\mathbb{P}_{\rm (v)}}
\newcommand{\pbbvi}{\mathbb{P}_{\rm (vi)}}
\newcommand{\pbbvii}{\mathbb{P}_{\rm (vii)}}
\newcommand{\pbbviii}{\mathbb{P}_{\rm (viii)}}
\newcommand{\wpbbi}{\widetilde{\mathbb{P}}_{\rm (i)}}
\newcommand{\wpbbii}{\widetilde{\mathbb{P}}_{\rm (ii)}}
\newcommand{\wpbbiii}{\widetilde{\mathbb{P}}_{\rm (iii)}}
\newcommand{\cffF}{\mathcal{F}}
\newcommand{\cffH}{\mathcal{H}}
\newcommand{\cffE}{\mathcal{E}}
\newcommand{\cffHt}{\widetilde{\mathcal{H}}}
\newcommand{\cffEt}{\widetilde{\mathcal{E}}}
\newcommand{\cfff}{\mathfrak{F}}
\newcommand{\cffh}{\mathfrak{H}}
\newcommand{\cffe}{\mathfrak{E}}
\newcommand{\cffht}{\widetilde{\mathfrak{H}}}
\newcommand{\cffet}{\widetilde{\mathfrak{E}}}
\newcommand{\bq}{\bar{q}}
\newcommand{\bp}{\bar{p}}
\newcommand{\tmin}{t_0}
\newcommand{\tmax}{t_1}
\newcommand{\vk}{\varkappa}
\newcommand{\xidvcs}{\xi_{\rm DVCS}}
\newcommand{\xitcs}{\xi_{\rm TCS}}
\newcommand{\dota}{{\dot{a}}}
\newcommand{\dotb}{{\dot{b}}}
\newcommand{\dotc}{{\dot{c}}}
\newcommand{\dotd}{{\dot{d}}}
\newcommand{\Op}{\mathcal{O}}
\newcommand{\lop}{\mathscr{O}}
\newcommand{\intba}{\iint_{\mathbb{D}} d\beta d\alpha\ }
\newcommand{\PhiPlusF}{\Phi^{(+)}_f(\beta,\alpha,t)}
\newcommand{\PhiPlus}{\Phi^{(+)}(\beta,\alpha,t)}
\newcommand{\phiplus}{\Phi^{(+)}}
\newcommand{\HplusFx}{H^{(+)}_f(x,\xi,t)}
\newcommand{\HplusX}{H^{(+)}(x,\xi,t)}
\newcommand{\HplusXp}{H^{(+)}(x',\xi,t)}
\newcommand{\hplus}{H^{(+)}}
\newcommand{\HplusFrho}{H^{(+)}_f(\rho,\xi,t)}
\newcommand{\HplusRho}{H^{(+)}(\rho,\xi,t)}
\newcommand{\deltaxba}{\delta(x-\beta-\alpha\xi)}
\newcommand{\thetaxba}{\theta(x-\beta-\alpha\xi)}
\newcommand{\deltaxPrimeba}{\delta(x'-\beta-\alpha\xi)}
\newcommand{\inv}{\mathcal{I}}
\newcommand{\hatx}{\hat{x}}
\newcommand{\hatz}{\hat{z}}
\newcommand{\bfDelta}{\boldsymbol{\hat{\ell}}}
\newcommand{\proj}{\mathscr{P}}
\newcommand{\wOp}{\widetilde{\Op}}
\newcommand{\wDelta}{\widetilde{\ell}}
\newcommand{\repr}{\mathcal{R}}
\newcommand{\Opimu}{\Op^i_{(\repr)}}
\newcommand{\Opjmu}{\Op^j_{(\repr)}}
\newcommand{\Opkmu}{\Op^k_{(\repr)}}
\newcommand{\scale}{\mathbb{Q}}
\newcommand{\lio}{\ell_{1,\,0}}
\newcommand{\liv}{\ell_{1,\,v}}
\newcommand{\lbuv}{\ell_{\bar{u},\,v}}
\newcommand{\bu}{{\bar{u}}}
\newcommand{\bv}{{\bar{v}}}
\newcommand{\bw}{{\bar{w}}}
\newcommand{\btau}{{\bar{\tau}}}
\newcommand{\dx}{\partial_{x}}
\newcommand{\dxp}{\partial_{x'}}
\newcommand{\dxi}{\partial_{\xi}}
\newcommand{\ycal}{\mathcal{Y}}

\newcommand{\bQ}{\bar{Q}} 

\newcommand\normord[1]{{:}\mkern1mu#1\mkern1.6mu{:}} 

\newcommand\lt[1]{{\left[#1\right]_{\rm LT}}} 

\newcommand\Li[2]{{\textrm{Li}_{#1}\left(#2\right)}}  
\newcommand\Ln[1]{{\,\textrm{ln}\left(#1\right)}}  

\begin{abstract}
We calculate the kinematic twist-3 and 4 corrections to the leading order amplitude of timelike Compton scattering (TCS) on a (pseudo-)scalar target, in the recently developed framework based on the conformal operator-product expansion. This allows us to compute the complete set of helicity amplitudes of the process, in particular those that vanish at leading twist. We compare the effects of higher twist contributions to TCS with those in deeply virtual Compton scattering (DVCS). Our estimates, based on a $\pi$-meson GPD model, indicate that these contributions are sizeable and will play a crucial role in the interpretation of data from current and forthcoming experiments.
\end{abstract}

\maketitle

\input{sec_intro}

\input{sec_COPE}

\input{sec_amplitudes}

\input{sec_numerics}
\input{sec_conclusions}
\\
\input{acknowledgements}

\bibliography{bibliography}

\appendix
\input{appendices/AppendixDD}

\input{appendices/DDsToGPDs}
\input{appendices/A00}

\end{document}

%% file: sec_intro.tex
\section{Introduction}
\label{sec:intro}
    
The parton tomography of hadrons is one of the primary goals of quantum chromodynamics (QCD). It is studied through the factorization properties of amplitudes for various exclusive processes, which provide access to generalized parton distributions (GPDs)~\cite{Muller:1994ses, Ji:1996nm, Diehl:2002he, Belitsky:2005qn}. Since exclusive processes have quite low cross sections, they are typically measured in high-luminosity photo- and lepto-production experiments at moderate value of a hard scale (denoted as $\scale$). Interpreting data collected under such conditions requires applying various corrections to leading-order QCD calculations. This is particularly important for describing the hadron's structure in impact parameter space~\cite{Burkardt:2002hr,Ralston:2001xs,Diehl:2002he}, which involves a Fourier transform from transverse momentum transfer to the hadron ($\bf \Delta_\perp$) to transverse position of the active parton inside the hadron ($\bf b_\perp$). Since such a transform cannot be reliably done if the interval in $|\bf \Delta_\perp|$ is too small, it raises the question of estimating corrections proportional to powers of $\sqrt{|t|}/\scale$, where $t=\Delta^2$ is the squared momentum transfer to the hadron. Another type of corrections relevant at moderate scales are target mass ($M$) corrections proportional to powers of $M/\scale$.

The aforementioned types of corrections to the scattering amplitudes are known as kinematic twist corrections. For a specific class of processes, works by Braun, Manashov and coworkers~\cite{Braun:2012bg, Braun:2014sta,Braun:2016qlg,Braun:2020zjm,Braun:2022qly} have demonstrated that these corrections can be conveniently calculated thanks to the conformal operator-product expansion, taking profit of the conformal symmetry of lowest order QCD. This method was originally applied to estimate kinematic twist corrections to deeply virtual Compton scattering (DVCS)~\cite{Muller:1994ses,Ji:1996nm,Radyushkin:1998bz}. A similar study was recently performed for a related class of processes involving generalized distribution amplitudes (GDAs)~\cite{Lorce:2022tiq,Pire:2023kng,Pire:2023ztb}. 

In this work, we follow the strategy based on the conformal operator-product expansion to calculate kinematic twist corrections in the general framework of two-photon scattering, where all-order factorization has recently been proven at next to leading power \cite{Schoenleber:2024ihr}. The adopted approach allows us, for the first time, to calculate these corrections for timelike Compton scattering (TCS)~\cite{Berger:2001xd} and to compare them with corresponding results previously obtained for DVCS. We note, that in addition to the kinematic twist corrections, there are genuine twist effects, which present a complex and separate task that we do not address here. For a discussion of genuine twists effects, see e.g. Refs.~\cite{Anikin:2000em,Kivel:2000fg,Radyushkin:2000ap, Aslan:2018zzk}.

The general framework of two-photon scattering can be defined with the help of double deeply virtual Compton scattering (DDVCS)~\cite{Belitsky:2003fj,guidal2003,Deja:2023ahc}:
\begin{equation}
        \gamma^{(*)}(q)+N(p)\to \gamma^{(*)}(q')+N(p')\,,
\end{equation}
where $N$ stands for a hadron, and symbols in parentheses denote four-momenta of particles. For DVCS the incoming photon is virtual, while the outgoing one is real ($-q^2=Q^2>0$, $q'^2=Q'^2=0$); the other way around for TCS ($Q^2=0$, $Q'^2>0$); and for DDVCS both photons are virtual ($Q^2>0$,  $Q'^2>0$). This way, DVCS and TCS can be considered as limiting cases of DDVCS. Feasibility studies for this process in the context of different experiments can be found in Ref.~\cite{Deja:2023ahc} and, more recently, in Ref.~\cite{Alvarado:2025huq}.

\begin{figure}
    \centering
    \includegraphics[width=0.22\textwidth]{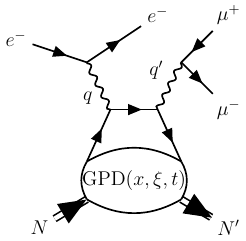}
    \includegraphics[width=0.22\textwidth]{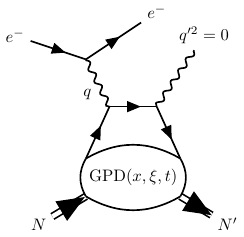}
    \includegraphics[width=0.22\textwidth]{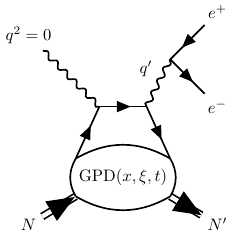}
    \caption{From left to right: DDVCS for muon pair production channel, DVCS and TCS processes.}
    \label{fig::ddvcs_dvcs_tcs}
\end{figure}

DVCS and DDVCS can be measured in lepto-production experiments, while TCS requires a (quasi-)real photon beam (see Fig.~\ref{fig::ddvcs_dvcs_tcs}). In the Bjorken limit, which assumes an infinite hard scale ($\scale^2 \to \infty$, related to the virtualities of photons) and finite skewness (defined below in Eq.~\eqref{xidef}), the amplitudes of all three processes can be factorized into perturbatively calculable coefficient functions and non-perturbative GPDs. For a spin-0 target, the chiral-even GPD associated to a quark of flavor $f$ is given by \cite{diehl_review}:
\begin{equation}
    H_f(x,\xi,t) = \frac{1}{2}\int\frac{d\lambda}{2\pi}\ e^{ix(\bp z)}\langle p'|\bar{\quark}_f(-z/2)\slashed{n}\quark_f(z/2)|p\rangle\big|_{z=\lambda n}\,,\quad \lambda\in\R\,,\quad n^2=0\,, \label{GPD_def}
\end{equation}
where $x$ is the average parton momentum, $\xi$ is the skewness, $t$ is the usual Mandelstam variable, $\bar{p} = (p+p')/2$, and the Wilson line is omitted for simplicity. We note that chiral-odd leading twist quark GPDs do not contribute to Compton processes, while gluon GPDs do not contribute at leading-order (LO) in the strong coupling constant, in which we are interested in this work.

Relaxing the conditions of the Bjorken limit is equivalent to accounting for the contributions to the Compton tensor that are coming from spacetime points slightly away from the light-cone ($z^2\neq 0$). Since the scale is no longer  infinite, corrections to the light-cone contribution to the Compton tensor translate to kinematic power corrections of the form
\begin{equation}\label{introTauKin}
    \left( \frac{\Lambda^2}{\scale^2} \right)^{(\tau_\textrm{kin}-2)/2} < 1 \,,\quad \tau_\textrm{kin}\geq 2\,,
\end{equation}
where typically $\Lambda^2 \in \{|t|, M^2\}$.

The goal of our research is to provide the theoretical and phenomenological description of TCS amplitudes at LO including the kinematic twist-3 and twist-4 corrections, off a spin-0 target. This involves considering kinematic effects entering the amplitude with at most a damping factor of the form $|t|/\scale^2$ or $M^2/\scale^2$. For this purpose, we will consider the general framework of DDVCS and perform the twist expansion for this process, obtaining DVCS and TCS as the limits of small outgoing and incoming virtuality, respectively. Comparison with previous DVCS results~\cite{Braun:2012bg, Braun:2022qly} will serve as a test of the correctness of our TCS results through the DDVCS formulation.

Our interest in spin-0 targets is twofold. First, they represent the simplest case, as they can be described by a single chiral-even GPD type. Second, among spin-0 targets, we find helium-4 and pions, both of which are experimentally accessible (the pion through the Sullivan process~\cite{sullivanProcess}). In addition, we note that the first experimental observation of TCS~\cite{CLAS:2021lky}, together with previous phenomenological studies~\cite{Grocholski:2019pqj}, has demonstrated the usefulness and complementarity of DVCS and TCS studies~\cite{Mueller:2012sma}. It is thus deemed appropriate to improve the theoretical precision of TCS, as it will be studied with great care in the near future at both fixed-target facilities~\cite{Chen:2014psa, Camsonne:2017yux, Zhao:2021zsm, Accardi:2023chb} and electron-ion colliders~\cite{AbdulKhalek:2021gbh, Anderle:2021wcy}. 

This article is organized as follows: in Sect.~\ref{sec:kinematics} we detail the momenta parametrization by means of light-cone coordinates. To make this article self-contained, in Sect.~\ref{sec::cope}, we summarize the main results of the conformal methods previously discussed in~\cite{Braun:2020zjm, Braun:2022qly}. In Sect.~\ref{sec:amplitudes}, we parameterize the Compton tensor with helicity-dependent amplitudes for general two-photon processes involving a spin-0 hadron, identify the minimal number of independent amplitudes allowed by parity conservation, and calculate their kinematic contributions at twist-3 and twist-4 levels. In the following Sect.~\ref{sec:numerics}, we present the numerical estimates for the helicity amplitudes of TCS on a pion target, implementing the GPD model of Ref.~\cite{Chavez:2021llq}. The article ends with remarks and conclusions in Sect.~\ref{sec:conclusions}. The article is complemented by three appendices. In Apps.~\ref{app::DDs} and~\ref{app::DDs->GPDs}, we provide formulas for mapping double distributions (DDs) onto GPDs and vice versa, which allows the use of the formulation presented in Sect.~\ref{sec:amplitudes} for modeling DDs instead of GPDs. From a phenomenological point of view, this is an attractive approach, as it allows to evaluate the amplitudes directly from DDs. In App.~\ref{Appendix-A00}, we present our results for the $\amp^{00}$ amplitude, which only exists in the DDVCS case. We note that some elements of this study have been presented at conferences~\cite{Martinez-Fernandez:2024pzm} and in the Ph.D. dissertation~\cite{vthesis}, providing a more pedagogical explanation of some difficult calculation steps presented in this article.

\section{Kinematics}
\label{sec:kinematics}

In two-photon processes, information on the interaction with a hadron is concealed in the non-forward matrix elements of the Compton tensor, given by the time-ordering of two currents:
\begin{equation}
    T^{\mu\nu} = i\int d^4z\ e^{iq'z}\langle p'| \mathbb{T}\{j^\mu(z)j^\nu(0)\} |p\rangle\,,
\end{equation}
with $j_\mu(z) = \sum_f (e_f/e) \bar{\quark}_f(z) \gamma_\mu \quark_f(z)$, $f$ the flavor index and $e_f$ the electric charge of quark $f$.

In light-cone coordinates, any four-vector $v$ may be written as $v^\mu=v^+ n'^\mu+v^- n^\mu + v_\perp^\mu$, where $n$ and $n'$ span the longitudinal plane and are lightlike vectors satisfying $nn'\neq 0$ and $n v_\perp=n' v_\perp=0$. 

In this manuscript, defining
\begin{equation}
    \bp=\frac{p+p'}{2}\quad\textrm{and}\quad \bq=\frac{q+q'}{2}\,,
\end{equation}
we consider a longitudinal plane spanned by $q$ and $q'$, this is:
\begin{equation}\label{qqPrime}
    q^\mu = \frac{1}{2R}n^\mu + \frac{RQ^2}{\Delta q'}n'^\mu\,,\quad q'^\mu = \frac{1}{2Q^2}\left( 1 - \frac{qq'}{R} \right)n^\mu - R\frac{R+qq'}{\Delta q'}n'^\mu\,,
\end{equation}
upon
\begin{equation}
    nn' = -\Delta q' = \frac{t + Q^2 + Q'^2}{2} = \frac{\scale^2}{2}\quad\mbox{and}\quad R = \sqrt{(qq')^2 + Q^2Q'^2}\,.
\end{equation}
We have chosen $nn'>0$ and proportional to the energy scale of DDVCS: 
\begin{equation}
 \scale^2=Q^2+Q'^2+t\,.   
\end{equation} 
This choice for $\scale^2$ is not imposed but drawn from the structure of the integrals which gives rise to the twist corrections. This will be discussed in detailed in Sect.~\ref{sec::A++}.

The choice (\ref{qqPrime}) for $q$ and $q'$ renders finite vectors in the DVCS and TCS limits, i.e.~$Q'^2\to 0$ and $Q^2\to 0$ respectively. 
We are left with determining $\Delta$ and $\bp$. The first one is straightforward:
\begin{equation}
    \Delta^\mu = q^\mu - q'^\mu = -\frac{1}{2Q^2}\left( 1 - \frac{Q^2+qq'}{R} \right) n^\mu + \frac{R}{\Delta q'}\left( Q^2 + R + qq' \right)n'^\mu\,,
\end{equation}
while the latter is given by conditions
\begin{equation}
    \Delta\bp = 0\,,\quad \xi = -\Delta n/(2\bp n)\,,\quad \bp^{\,2} = M^2 - t/4\,,
\label{xidef}
\end{equation}
as
\begin{equation}
    \bp^{\,\mu} = -\frac{R - Q^2 - qq'}{4\xi R Q^2} n^\mu - \frac{R(R+Q^2+qq')}{2\xi(\Delta q')}n'^\mu + \bp^{\,\mu}_\perp
    = \frac{\Delta^-}{2\xi}n^\mu - \frac{\Delta^+}{2\xi}n'^\mu + \bp^{\,\mu}_\perp\,.
\end{equation}
Taking into account that $-\bp_\perp^{\,2} = |\bp_\perp|^2 \geq 0$, one can get the value of $t$ for which its modulus is minimal, namely $\tmin$, for a fixed value of the skewness:
\begin{equation}\label{tminFromBPperp}
    \tmin = -\frac{4M^2\xi^2}{1-\xi^2}\qquad (|t| \geq |\tmin|)\,.
\end{equation}
In turn, the skewness can be expressed through invariants:
\begin{align}\label{xiDDVCSphaseSpace}
    \xi  = \frac{2R}{2Q^2/x_B - Q^2 - Q'^2 + t}
     = \frac{\sqrt{ (Q^2+Q'^2)^2 + t^2 + 2t(Q^2-Q'^2) }}{2Q^2/x_B - Q^2 - Q'^2 + t}\,,
\end{align}
where $x_B=Q^2/(2pq)$ is the usual Bjorken variable. Considering the DVCS limit ($Q'^2\to 0$), this expression for the skewness takes the form given in Ref.~\cite{Braun:2012bg}.

%% file: sec_COPE.tex
\section{Conformal operator-product expansion \`a la Braun-Ji-Manashov}\label{sec::cope}

We are interested in corrections to the LT amplitudes described by means of GPDs, which are matrix elements of geometric twist-2 operators. Therefore, we consider an expansion of the time-ordering of the two currents in the Compton tensor in a basis of geometric twist-2 operators and their descendants, given by derivatives of the former ones. These operators would eventually be related to GPDs and derivatives of those, giving rise to new convolutions that enter the amplitude with the suppression factors discussed in Eq.~\eqref{introTauKin}. 

The methods detailing how to deal with QCD as a conformal field theory \cite{difrancesco,ferrara1972,ferraraparisi1972,ferraragrilloparisigatto1972} in this context were developed in Refs.~\cite{Braun:2020zjm,Braun:2022qly} by V.~M.~Braun et al. and they constitute the basis of the work presented here. To ensure the article is self-contained and to establish the conventions used, we summarize in this section some original results that are relevant to our calculation.

At leading order in the strong coupling constant, the Lagrangian of QCD in four dimensions is conformally invariant, up to the gauge-fixing and Faddeev-Popov ghosts terms. For matrix elements of gauge-invariant operators, these terms are irrelevant, validating the application of conformal techniques for them. The Compton tensor is one of this kind and so, for its vector part which is the only one needed for a spin-0 target, it takes the form~\cite{Braun:2020zjm,Braun:2022qly}:
\begingroup
\allowdisplaybreaks{
\begin{align}\label{ComptonCOPE_vectorPart}
    T^{\mu\nu} = &\ i\int d^4z\ e^{iq'z}\langle p'| \mathbb{T}\{j^\mu(z)j^\nu(0)\} |p\rangle\nonumber\\
    = &\ \frac{1}{i\pi^2}i\int d^4z\ e^{iq'z} \nonumber\\
    & \times \Bigg\{ \frac{1}{(-z^2+i0)^2}\left[ 
    g^{\mu\nu}\lop(1,0) - z^\mu\partial^\nu\int_0^1 du\ \lop(\bu, 0) - z^\nu(\partial^\mu - i\Delta^\mu)\int_0^1 dv\ \lop(1,v) \right] \nonumber\\
    & - \frac{1}{-z^2+i0}\left[ \frac{i}{2}(\Delta^\nu\partial^\mu - \Delta^\mu\partial^\nu)\int_0^1 du\int_0^{\bu} dv\ \lop(\bu,v) - \frac{t}{4}z^\mu\partial^\nu\int_0^1du\ u\int_0^\bu dv\ \lop(\bu,v) \right] \nonumber\\
    & + \frac{t}{2}\frac{z^\mu z^\nu}{(-z^2+i0)^2}\int_0^1 du\ \bu\int_0^\bu dv\ \lop(\bu,v) \nonumber\\
    & + \frac{g^{\mu\nu}}{4(-z^2+i0)}\left[ \int_0^1 du\int_0^\bu dv\ \lop_1(\bu,v) - \int_0^1 dv\ \lop_2(1,v) \right] \nonumber\\
    & +\frac{1}{4(-z^2+i0)}( z^\mu\partial^\nu + z^\nu\partial^\mu - iz^\mu\Delta^\nu )\int_0^1 du\int_0^\bu dv\ \left( (\ln \btau)\lop_1(\bu,v) + \frac{v}{\bv}\lop_2(\bu,v) \right) \nonumber\\
    & +\frac{1}{2(-z^2+i0)}( -z^\mu\partial^\nu + z^\nu\partial^\mu + iz^\mu\Delta^\nu )\int_0^1 du\int_0^\bu dv\ \frac{\tau}{\btau}\left( -\lop_1(\bu,v) + \frac{\bu}{u}\lop_2(\bu,v) \right) \nonumber\\
    & +\frac{1}{4(-z^2+i0)}z^\nu(\partial^\mu - i\Delta^\mu) \int_0^1 du\int_0^\bu dv\ \frac{v}{\bv}\left[ -2\left( 1+\frac{2\tau}{\btau} \right)\lop_1(\bu,v) + \frac{v}{\bv}\lop_2(\bu,v) \right]  \nonumber\\
    & + \frac{1}{4(-z^2+i0)}z^\nu(\partial^\mu - i\Delta^\mu) \int_0^1 dv\ \frac{v}{\bv}\lop_2(1,v) \nonumber\\
    & +\frac{1}{2(-z^2+i0)}z^\mu\partial^\nu\int_0^1 du\int_0^\bu dv\ \left[(\ln\bu + u)\lop_1(\bu,v) + \bu\lop_2(\bu,v)\right] \nonumber\\
    & - \frac{1}{4(-z^2+i0)}z^\mu\partial^\nu\int_0^1 du\int_0^\bu dv\ 
    \left(1+\frac{4\tau}{\btau}\right)\lop_2(\bu,v)  \nonumber\\
    & - \frac{z^\mu z^\nu}{(-z^2+i0)^2}\int_0^1 du\int_0^\bu dv\ \left[ (\ln\btau + \ln\bu + u)\lop_1(\bu,v) + \left(\frac{v}{\bv}+\bu\right)\lop_2(\bu,v) \right] \nonumber\\
    & + \frac{z^\mu z^\nu}{4(-z^2+i0)}\left[ i\Delta\partial + \frac{t}{2} \right]\int_0^1 du\int_0^\bu dv\ \frac{v}{\bv}\left(\frac{2}{\btau} - 1\right)\lop_1(\bu,v) \nonumber\\
    &  - \frac{z^\mu z^\nu}{2(-z^2+i0)}\left[ i\Delta\partial + \frac{t}{4} \right]\int_0^1 du\int_0^\bu dv\ \left( \ln\btau + \frac{2\tau}{\btau} \right)\lop_1(\bu,v)   \Bigg\} \,,
\end{align}
}
\endgroup
where
\begin{equation}
    \tau = \frac{u v}{\bu\bv}\,,\quad \bar{\tau} = 1-\tau \,,\quad \bu = 1-u\,,\quad \bv = 1-v\,,
\end{equation}
while the operators $\lop$, $\lop_1$ and $\lop_2$ must be understood as matrix elements $\langle p'|\lop|p\rangle$, $\langle p'|\lop_1 |p\rangle$, $\langle p'|\lop_2 |p\rangle$\,, respectively, defining
\begin{align}
    \lop_1(\lambda_1,\lambda_2) & = (i\Delta\partial_z)\lop(\lambda_1,\lambda_2)\,, \label{lop_1_def} \\
    \lop_2(\lambda_1,\lambda_2) & = \lop_1(\lambda_1,\lambda_2) + \frac{t}{2}\lop(\lambda_1,\lambda_2)\,. \label{lop_2_def}
\end{align}
by means of the operator $\lop$ (for which $z^2\neq 0)$:
\begin{equation}\label{lop_charges}
    \lop(\lambda_1, \lambda_2) = \sum_f \left(\frac{e_f}{e}\right)^2\frac{1}{2}\lt{\bar{\quark}_f(\lambda_1 z)\slashed{z}\quark_f(\lambda_2 z) - (\lambda_1\leftrightarrow \lambda_2)}\,.
\end{equation}
Here, $\lt{\ }$ represents the geometric leading twist projection of the corresponding operator, to be detailed later. 

In the spin-0 case, we can extract the quark correlator from the definition of GPD, given by Eq. (\ref{GPD_def}), through the Fourier transform on the variable $x$:
\begin{equation}
    \langle p'| \bar{\quark}_f(z_2)\slashed{n}\quark_f(z_1) |p\rangle = 2(\bp n)\int_{-1}^1 dx\ e^{-i(\bp n)\left[ \lambda_1(\xi+x) + \lambda_2(\xi-x) \right]} H_f(x,\xi,t)\,,\quad z_i = \lambda_i n\,.
\end{equation}
From this formula one notices that the exchange $z_1\leftrightarrow z_2$ (resulting in $\lambda_1\leftrightarrow\lambda_2$) is equivalent to the switch $x\rightarrow -x$\,. Therefore, we may consider the following non-local operator evaluated at light-cone positions:
\begin{equation}\label{Op_f}
    \Op_f(\lambda_2, \lambda_1) = \frac{1}{2}\bar{\quark}_f(\lambda_2 n)\slashed{n}\quark_f(\lambda_1 n) - (\lambda_2\leftrightarrow \lambda_1)\,,
\end{equation}
assuming Wilson lines implicitly. Its matrix element between states of momenta $p$ and $p'$ reads: 
\begin{equation}
    \langle p'| \Op_f(\lambda_2, \lambda_1) |p\rangle = 2(\bp n)\int_{-1}^1 dx\ e^{-i(\bp n)\left[ \lambda_1(\xi+x) + \lambda_2(\xi-x) \right]} \frac{H_f^{(+)}(x,\xi,t)}{2}\,.
\end{equation}
where $H_f^{(+)}(x,\xi,t) = H_f(x,\xi,t) - H_f(-x,\xi,t)$ is the usual C-even part of the GPD.\footnote{We note that the GPD $H_q(x,\xi,t)$ used in  Eq.~(3.21) of \cite{Braun:2022qly} corresponds to one-half of the C-even combination.} 
Employing the double distribution representation of the light-cone operator:
\begin{equation}\label{corrOp_f}
    \langle p'| \Op_f(\lambda_1, \lambda_2) |p\rangle = \frac{2i}{\lambda_{12}}\intba e^{-i\ell_{\lambda_1,\lambda_2}n}\Phi_f^{(+)}(\beta,\alpha,t)\,,\quad z_i=\lambda_i n \,,
\end{equation}
allows us to express the GPD in the following way: 
\begin{equation}\label{dH+/dx}
    \frac{1}{2}\partial_x H_f^{(+)}(x,\xi,t) = \intba \delta(x-\beta-\alpha\xi)\Phi^{(+)}_f(\beta,\alpha,t)\,.
\end{equation}
We describe more technical details of the double distribution representation in the Appendix \ref{app::DDs}. For later reference, when dropping the flavor index $f$ we refer to the sum over flavors weigthed with the square of the fractional charges, i.e.~$\Phi^{(+)}=\sum_f (e_f/e)^2\Phi^{(+)}_f$ and likewise for $H^{(+)}$.

The key ingredient of the framework presented in \cite{Braun:2022qly}, allowing also for the expression of the operator $\lop$ (defined outside of the light-cone) in terms of DD, reads:
\begin{align}\label{corrlop_f}
    \langle p'| \lop_f(\lambda_1, \lambda_2) |p\rangle & = \frac{2i}{\lambda_{12}}\iint_{\mathbb{D}}d\beta d\alpha\ \lt{e^{-i\ell_{\lambda_1,\lambda_2}z}}\Phi_f^{(+)}(\beta,\alpha,t)\,,\quad \lambda_{12}=\lambda_1-\lambda_2\,.
\end{align}
The above (geometric) LT projection is described in detail in Ref.~\cite{geyer1999} and can be written as
\begin{equation}\label{expLT}
    \lt{e^{-i\ell_{\lambda_1,\lambda_2} z}} = e^{-i\ell_{\lambda_1,\lambda_2} z} + \sum_{k=1}^\infty \int_0^1 dw\ \left(\frac{z^2\ell_{\lambda_1,\lambda_2}^2}{4}\right)^k \frac{\bw^{k-1}w^k}{k!(k-1)!}e^{-iw(\ell_{\lambda_1,\lambda_2} z)}\,,\quad \bar{w}=1-w\,.
\end{equation}
Now it is manifest that the operators that partake of the OPE in Eq.~(\ref{ComptonCOPE_vectorPart}) can be expressed by means of DDs and, therefore, by GPDs.

%% file: sec_amplitudes.tex
\section{Helicity-dependent amplitudes}
\label{sec:amplitudes}
Consider two polarization vectors $\eps^\mu(A)$ for the incoming photon and $\eps'^\mu(B)$ for the outgoing, where $A, B$ labels the polarization: $0$ for longitudinal, $\pm 1$ for transverse polarization (in the plane of $\bp_\perp$). Then, we can defined the so-called helicity-dependent amplitudes~\cite{Belitsky:2012ch} as
\begin{equation}\label{h-amplitude}
    \amp^{AB} = (-1)^{A-1}(\eps'_\mu(B))^*\, T^{\mu\nu}\eps_\nu(A)\,.
\end{equation}
They represent the transition amplitude from an incoming photon with polarization $A$ to an outgoing one with polarization $B$. 

In the case of the opposite polarizations they are related by parity conservation:
\begin{equation}
    \amp^{AB}(\rho, \xi, t) = \amp^{-A\,-B}(\rho, \xi, t)\,,
\end{equation}
leaving a total of five independent amplitudes: $\amp^{++}$, $\amp^{+-}$, $\amp^{0+}$, $\amp^{+0}$ and $\amp^{00}$.

We would like to parameterize the Compton tensor by means of helicity amplitudes in order to find the corresponding projectors, which we would apply to the all-twists expansion of the Compton tensor given by Eq.~(\ref{ComptonCOPE_vectorPart}). This way, we can obtain an expression for the amplitudes $\amp^{AB}$ to the desired twist accuracy, which for us corresponds to the kinematic twist-4.

To parameterize the Compton tensor by means of helicity amplitudes, one can make use of the spin representation of the Lorentz group, cf.~\cite{srednicki}. This way, one can define two spinors, relate them to the lightlike vector $n$, $n'$, and make a parametric expansion of any tensor. For further details on this methodology and application to DVCS, see~\cite{Braun:2012bg}. An alternative approach was given by Tarrach in Ref.~\cite{Tarrach:1975tu}. With those methods and the polarization vectors given by:
\begin{align}
    \eps^\nu(0) & = \frac{R}{Q(\Delta q')}\left[ \frac{\Delta q'}{2R^2}n^\nu - Q^2 n'^\nu \right]\,,\label{eps0}\\
    \eps'^\nu(0) & = \frac{iQ}{2R}\sqrt{\frac{R-qq'}{R+qq'}} \left[ \frac{1}{Q^2}n^\nu + \frac{2R^2}{\Delta q'}\frac{R+qq'}{R-qq'}n'^\nu \right]\,.\label{eps'0}\\
    \eps^\nu(+) & = -\frac{\bp_\perp^{\,\nu} - i\widetilde{\bp}^{\,\nu}_\perp}{\sqrt{2}|\bp_\perp|}\,,\quad \widetilde{\bp}^{\,\nu}_\perp = \epsilon_\perp^{\nu\mu}\bp_\mu\,, \label{eps+}\\
    \eps^\nu(-) & = -\frac{\bp_\perp^{\,\nu} + i\widetilde{\bp}^{\,\nu}_\perp}{\sqrt{2}|\bp_\perp|}\,,\label{eps-}
\end{align}
the result for the Compton tensor for a spin-0 target (valid for DDVCS, DVCS and TCS) reads:
\begin{align}\label{Tvector-h-amplitudes2}
    T^{\mu\nu} =  
    &  - \amp^{++}g_\perp^{\mu\nu} + \amp^{+-}\frac{1}{|\bp_\perp|^2}\left[ \bp^{\,\mu}_\perp\bp^{\,\nu}_\perp - \widetilde{\bp}^{\,\mu}_\perp\widetilde{\bp}^{\,\nu}_\perp \right] \nonumber\\
    &+\ \amp^{00} \frac{-i}{QQ'R^2}\left[ (qq')(Q'^2q^\mu q^\nu - Q^2 q'^\mu q'^\nu) + Q^2 Q'^2 q^\mu q'^\nu - (qq')^2 q'^\mu q^\nu \right] \nonumber\\
    & + \amp^{+0}\frac{i\sqrt{2}}{R|\bp_\perp|}\left[ Q'q^\mu - \frac{qq'}{Q'}q'^\mu \right]\bp^{\,\nu}_\perp -\amp^{0+}\frac{\sqrt{2}}{R|\bp_\perp|}\bp^{\,\mu}_\perp\left[ \frac{qq'}{Q}q^\nu + Q q'^\nu \right] 
    \,.
\end{align}
In the LT LO approximation, only the first term survives, and $\amp^{++}$ reduces to the well-known LO LT Compton form factor $\mathcal{H}$ \cite{Belitsky:2003fj}.

\subsection{Projectors onto helicity amplitudes}\label{sec::projectors-h}

Close inspection of expression (\ref{Tvector-h-amplitudes2}) allows to define a set of helicity projectors $\Pi^{(AB)}_{\mu\nu}$ such that
\begin{equation}\label{PiT=amp}
    \Pi^{(AB)}_{\mu\nu} T^{\mu\nu} = \amp^{AB}\,.
\end{equation}
Considering $\bp_{\perp,\,\alpha}q^\alpha = \bp_{\perp,\,\alpha}q'^\alpha = 0$\,, one can define
\begin{align}
    \Pi^{(0+)}_{\mu\nu} & = \bp_{\perp,\,\mu}q'_\nu\frac{R}{\sqrt{2}|\bp_\perp|}\left[ QQ'^2 + \frac{(qq')^2}{Q} \right]^{-1} =  \bp_{\perp,\,\mu}q'_\nu\frac{Q}{\sqrt{2}|\bp_\perp|R} \,, \label{Pi0+} \\
    \Pi^{(+0)}_{\mu\nu} & = -q_\mu\bp_{\perp,\,\nu}\frac{iR}{\sqrt{2}|\bp_\perp|}\left[ Q'Q^2 + \frac{(qq')^2}{Q'} \right]^{-1} = -q_\mu\bp_{\perp,\,\nu}\frac{iQ'}{\sqrt{2}|\bp_\perp|R} \,. \label{Pi+0}
\end{align}
Using $\widetilde{\bp}_\perp^{\,2} = \bp_\perp^{\,2} = -|\bp_\perp|^2$\,:
\begin{equation}\label{Pi++}
    \Pi^{(++)}_{\mu\nu} = \frac{1}{2|\bp_\perp|^2}\left( \bp_{\perp,\mu}\bp_{\perp,\nu} + \widetilde{\bp}_{\perp,\mu}\widetilde{\bp}_{\perp,\nu} \right) = -\frac{g_{\perp,\,\mu\nu}}{2}\,,
\end{equation}
and with $\bp_\perp\cdot\widetilde{\bp}_\perp = \bp_{\perp,\,\mu}\epsilon^{\mu\nu}_\perp\bp_{\perp,\,\nu} = 0$:
\begin{equation}\label{Pi+-}
    \Pi^{(+-)}_{\mu\nu} = \frac{1}{2|\bp_\perp|^2}\left( \bp_{\perp,\,\mu}\bp_{\perp,\,\nu} - \widetilde{\bp}_{\perp,\,\mu}\widetilde{\bp}_{\perp,\,\nu} \right)\,.
\end{equation}
Finally, for amplitude $\amp^{00}$ we can only use a projector made out of longitudinal vectors, i.e.~$q, q'$. We can make it antisymmetric to obtain the simplest of the projectors. This way,
\begin{equation}\label{Pi00}
    \Pi^{(00)}_{\mu\nu} = -\frac{i2QQ'}{R^2}q_{[\mu}q'_{\nu]}\,,
\end{equation}
where $q_{[\mu}q'_{\nu]} = (q_\mu q'_\nu - q_\nu q'_\mu)/2$.

In what follows, we will use this projectors to compute the different helicity amplitudes from the Compton tensor in Eq.~(\ref{ComptonCOPE_vectorPart}) of a spin-0 target. Taking the appropriate small virtuality limit, from the DDVCS results we will obtain the DVCS (already published in \cite{Braun:2022qly}) and the TCS ones (to the best of our knowledge, firstly published here).

\subsection{Transverse-helicity conserving amplitude, $\amp^{++}$}\label{sec::A++}

In the current section we present the calculation of the transverse helicity-conserving amplitude $\amp^{++}$ up to kinematic twist-4 accuracy. To achieve this, we apply the projector $\Pi^{(++)}_{\mu\nu}$ from Eq.~(\ref{Pi++}) to the Compton tensor decomposition $T^{\mu\nu}$ in Eq.~(\ref{ComptonCOPE_vectorPart}). All antysymmetric terms, as well as those proportional to $q,q'$, or $\Delta$, are washed out.

Amplitude $\amp^{++}$ will be the only one with a contribution at kinematic leading-twist (twist-2) as its corresponding Lorentz structure is the transverse metric,~$g_\perp^{\mu\nu}$\,, vid. Eq.~(\ref{Tvector-h-amplitudes2}). The structure of the LT contribution allows us to define a new invariant related to the {\it generalized Bjorken variable} previously introduced by Belitsky and M\"uller in Ref.~\cite{Belitsky:2003fj}.

\subsubsection{LT component, energy scale of DDVCS and generalized Bjorken variable}\label{sec::LTcomponent}

The leading twist contribution to the amplitude of DDVCS originates from the most singular term ($\sim 1/z^4$), which appears in the first line of Eq.~(\ref{ComptonCOPE_vectorPart}):
\begin{align}\label{A++_1stLine}
    \left.\amp^{++}\right|_{\rm LT} \subset & -\frac{g_{\perp,\,\mu\nu}}{2i\pi^2}i\int d^4z\ e^{iq'z} \frac{1}{(-z^2+i0)^2}\Bigg[ 
    g^{\mu\nu}\lop(1,0) - z^\nu\partial^\mu\int_0^1 du\ \lop(\bu, 0) - z^\mu\partial^\nu\int_0^1 dv\ \lop(1,v) \Bigg]\,.
\end{align}
Here, the symbol $\subset$ (read as ``contained in'') reflects the fact that the RHS includes not only the LT component of the amplitude but higher-order corrections too.

The application of the expression (\ref{corrlop_f}) results in the following Fourier transforms:
\begin{equation}\label{fourier_for_LT}
    i\int d^4z\ e^{iq'z}\frac{\lt{e^{-i\ell z}}}{(-z^2+i0)^2}\,,\quad i\int d^4z\ e^{iq'z}\frac{z^\nu\partial^\mu \lt{e^{-i\ell z}}}{(-z^2+i0)^2}\,,
\end{equation}
where $\ell$ is a shorthand for the general
\begin{equation}
    \ell_{\lambda_1,\,\lambda_2} = -\lambda_1\Delta - \lambda_{12}\left[ 
\beta\bp - \frac{1}{2}(\alpha+1)\Delta \right]\,,
\end{equation}
and $\lt{\ }$ stands for the geometric LT projection given by Eq.~(\ref{expLT}). This projection renders a series of integrals of the form:
\begin{equation}\label{Inm}
    I_{n,m} = \int_0^1 dw\ \frac{w^n}{(aw^2+bw+c)^m}\,,
\end{equation}
where 
\begin{align}\label{abcOmega}
    a & = \ell^2 = \lambda_1^2 t + \lambda_{12}^2\left[ \beta^2\bp^{\,2} + \frac{1}{4}(\alpha+1)^2t \right] - \lambda_1\lambda_{12}(\alpha+1)t\,,\\
    b & = -2q'\ell = -2(q'\Delta)\left[ \lambda_{12}\frac{\beta}{2\Omega} + \frac{\lambda_{12}}{2}(\alpha+1) - \lambda_1 \right]\,,\\
    \Omega & = -\frac{\Delta q'}{2\bp q'} \,,\\
    c & = Q'^2+i0\,.
\end{align}
Taking into account that $\ell^2 = O(|t|, M^2)$, and $b\propto-2q'\Delta = Q^2+Q'^2+t$, we notice that the twist expansion shall be done in powers of 
\begin{equation}
    \frac{a}{b}=O(\textrm{tw-4})\,,\quad\textrm{whereas}\quad \frac{c}{b}=O(1)\,.
\end{equation}
Therefore, we identify the hard scale for a general Compton process to be 
\begin{equation}
    \scale^2=Q^2+Q'^2+t\,.
\end{equation}

To kinematic twist-4 accuracy, the Fourier transforms of Eq.~(\ref{fourier_for_LT}) read:
\begin{equation}\label{fourier_1/z4_tw4}
    i\int d^4z\ \frac{e^{iq'z}\lt{e^{-i\ell z}}}{(-z^2+i0)^2} = \underbrace{-\pi^2\ln\left(\frac{b+c}{-\mu^2}\right)}_{\mathclap{\textrm{LT term}}} - \pi^2\frac{a}{b+c} + \pi^2\frac{a}{b} + \pi^2\frac{ac}{b^2}\ln\left(\frac{c}{b+c}\right) + O(\textrm{tw-6})\,,
\end{equation}
where $\mu^2$ stands for the renormalization scale in dimensional regularization, and
\begin{align}\label{fourier_z^muD^nu/z4}
    i\int d^4z\ e^{iq'z}\frac{z^\nu\partial^\mu}{(-z^2+i0)^2}\lt{e^{-i\ell z}} =&\ g^{\mu\nu}\pi^2\left[ \frac{a}{b} + \frac{ac}{b^2}\ln\left( \frac{c}{b+c} \right) \right] - 2\pi^2\frac{\ell^\nu\ell^\mu}{b+c} + (\textrm{terms}\sim q') + O(\textrm{tw-6})\,.
\end{align}
The first logarithm in Eq.~(\ref{fourier_1/z4_tw4}) is the only LT component, hence:
\begin{align}
    \left.\amp^{++}\right|_{\rm LT} \subset & -\frac{g_{\perp,\,\mu\nu}}{2i\pi^2} 2i \intba \phiplus \left. g^{\mu\nu}\left[-\pi^2\ln\left(\frac{b+c}{-\mu^2}\right)\right]\right|_{\ell = \ell_{1,0}}\nonumber\\
    \subset & \intba 2\phiplus \left.\ln\left(\frac{b+c}{-\mu^2}\right)\right|_{\ell=\ell_{1,0}}\,, 
\end{align}
where $\phiplus = \PhiPlus$.

To map the function $\phiplus$ to the corresponding GPD $\hplus$ we need to introduce the identity $1 = \int_{-1}^1 dx\ \delta(x-\beta-\alpha\xi)$ in $\amp^{++}$. This identity leads to
\begin{align}
    \left.\amp^{++}\right|_{\rm LT} \subset & \int_{-1}^1 dx \intba 2\phiplus\delta(x-\beta-\alpha\xi) \left.\ln\left(\frac{b+c}{-\mu^2}\right)\right|_{\ell=\ell_{1,0}} \nonumber\\
    \subset & \int_{-1}^1 dx \intba 2\phiplus\delta(x-\beta-\alpha\xi) \nonumber\\
    & \times \ln\left( \frac{\scale^2}{-2\mu^2}\left[ \frac{x-\xi}{\xi} + \frac{2Q'^2}{\scale^2} + i0 \right]\left[ 1+\frac{\beta\left(\frac{1}{\Omega} - \frac{1}{\xi}\right)}{\frac{x-\xi}{\xi}+\frac{2Q'^2}{\scale^2}+i0} \right] \right)\,.
\end{align}

One can show that the term proportional to $\beta$ in the above equation is a twist-4 component. To the LT we can omit this factor. Due to the property (\ref{intbaPhi=0}), the result does not depend on the renormalization scale $\mu$. The structure of this solution motivates the definition of a new variable named {\it generalized Bjorken variable} and denoted as $\rho$:
\begin{equation}\label{actualRhoDef}
    \frac{x-\xi}{\xi} + \frac{2Q'^2}{\scale^2} = \frac{x-\rho}{\xi} \Rightarrow \rho = \xi \frac{qq'}{\Delta q'}\,,
\end{equation}
which can be expressed as follows:
\begin{align}
    \rho & = \xi\frac{Q^2 - Q'^2 + t}{Q^2+Q'^2+t} = \xi_{\textrm{\cite{Belitsky:2003fj}}} + \frac{\xi t}{2\scale^2} + O(\textrm{tw-6})\,.
\end{align}
The variable $\xi_{\textrm{\cite{Belitsky:2003fj}}}$ is the generalized Bjorken variable as introduced by Belitsky and M\"uller in Ref.~\cite{Belitsky:2003fj}. To avoid confusion with the skewness, we denote it as $\rho$ following notation in \cite{diehl_review}. To LT accuracy all definitions agree. 

Employing the definition of $\rho$, and making use of the properties of the DD (summarized in the Appendix \ref{app::DDs}), to kinematic twist-2 accuracy the amplitude reads:
\begin{align}\label{A++LT}
    \left.\amp^{++}\right|_{\rm LT} & = \int_{-1}^1 dx\intba 2\phiplus \delta(x-\beta-\alpha\xi) \ln\left(\frac{x-\rho}{\xi} + i0\right) \nonumber\\
    & = \int_{-1}^1 dx\ \left(\dx \hplus\right) \ln\left(\frac{x-\rho}{\xi} + i0\right) 
    = \int_{-1}^1 dx\ \frac{1}{\rho-x-i0}\HplusX\,,
\end{align}
which agrees with the usual LT approximation, see for example~\cite{Belitsky:2003fj,guidal2003,Deja:2023ahc}.

\subsubsection{Final result up to twist-4, and the DVCS and TCS limits}

To twist-4, we have contributions coming from the following integrals: 
\begin{align}\label{A++_onlyTw4}
    \left.\amp^{++}\right|_\textrm{tw-4} \supset & -\frac{g_{\perp,\,\mu\nu}}{2i\pi^2}i\int d^4z\ e^{iq'z}\Bigg\{ \frac{1}{(-z^2+i0)^2}\Bigg[ 
    g^{\mu\nu}\lop(1,0) - z^\nu\partial^\mu\int_0^1 du\ \lop(\bu, 0) - z^\mu\partial^\nu\int_0^1 dv\ \lop(1,v) \Bigg] \nonumber\\
    &  \qquad + \frac{t}{2}\frac{z^\nu z^\mu}{(-z^2+i0)^2}\int_0^1 du\ \bu\int_0^\bu dv\ \lop(\bu,v) \nonumber\\
    & \qquad + \frac{g^{\mu\nu}}{4(-z^2+i0)}\left[ \int_0^1 du\int_0^\bu dv\ \lop_1(\bu,v) - \int_0^1 dv\ \lop_2(1,v) \right] \nonumber\\
    & \qquad - \frac{z^\nu z^\mu}{(-z^2+i0)^2}\int_0^1 du\int_0^\bu dv\ \left[ (\ln\btau + \ln\bu + u)\lop_1(\bu,v) + \left(\frac{v}{\bv}+\bu\right)\lop_2(\bu,v) \right] \Bigg\}\,.
\end{align}
Solving the Fourier transforms first, expanding in powers of $a/b$ (taking into account that $c/b=O(1)$) afterwards, and integrating over $u$ and $v$, we finally obtain:
\begin{align}
    \amp^{++} &= \int_{-1}^1dx\ \Bigg\{ -\left(1-\frac{t}{2\scale^2}+\frac{t(\xi-\rho)}{\scale^2}\dxi\right)\frac{\hplus}{x-\rho+i0} \nonumber\\
    &\ + \frac{t}{\xi\scale^2}\Bigg[ \pbbi +  \pbbii - \frac{\xi L}{2}
    + \frac{\wpbbiii-\wpbbi}{2}  \nonumber\\
    &\ \phantom{+ \frac{t}{\xi\scale^2}\Bigg[} - \frac{\xi}{x+\xi}\Bigg( \Ln{\frac{x-\rho+i0}{\xi-\rho+i0}} - \frac{\xi+\rho}{2\xi}\Ln{\frac{-\xi-\rho+i0}{\xi-\rho+i0}} - \wpbbi \Bigg) \Bigg] \hplus \nonumber\\
    &\ -\frac{t}{\scale^2}\dxi\Bigg[ \Bigg( \pbbi+\pbbii - \frac{\xi L}{2}\nonumber\\
    &\ \phantom{-\frac{t}{\scale^2}\dxi\Bigg[} 
     - \frac{\xi}{x+\xi}\Bigg( \Ln{\frac{x-\rho+i0}{\xi-\rho+i0}} - \frac{\xi+\rho}{2\xi}\Ln{\frac{-\xi-\rho+i0}{\xi-\rho+i0}} - \wpbbi \Bigg) \Bigg)\hplus \Bigg] \nonumber\\
    &\ + \frac{\bp_\perp^2}{\scale^2}2\xi^3\dxi^2 \Bigg[ \Bigg( \pbbi+\pbbii- \frac{\xi L}{2}+\frac{\wpbbiii+\wpbbi}{2}  \Bigg)\hplus \Bigg] \Bigg\} \nonumber\\
    &\ + O(\textrm{tw-6})\,, \label{amp++_final}
\end{align}
where the following functions were defined:
\begin{align}\label{collectCoefficientsA++DDVCS}
    \pbbi(x/\xi,\rho/\xi) &=\frac{\xi-\rho}{x-\xi}\Li{2}{-\frac{x-\xi}{\xi-\rho+i0}}\,,\nonumber\\
    \wpbbi(x/\xi,\rho/\xi) &= -\frac{\xi-\rho}{x-\xi}\Ln{\frac{x-\rho+i0}{\xi-\rho+i0}}\,,\nonumber\\
    \pbbii(x/\xi,\rho/\xi) &= \frac{\xi-\rho}{x+\xi}\left[ \Li{2}{-\frac{x-\xi}{\xi-\rho+i0}} - (x\to -\xi) \right]\,,\nonumber\\
    \wpbbiii(x/\xi,\rho/\xi) &= -\frac{\xi+\rho}{x+\xi}\Ln{\frac{x-\rho+i0}{-\xi-\rho+i0}}\,,
\end{align}

\begin{equation}\label{Ldef}
    L(x,\xi,\rho) = \int_0^1dw\ \frac{-4}{x-\xi-w(x+\xi)}\int_0^1du\ \Ln{1 + \frac{\bu(x-\xi-w(x+\xi))}{\xi-\rho+i0}}\left[\Ln{\frac{\bu(1-w)}{1-\bu w}} + \frac{1}{1-\bu w}\right]\,.
\end{equation}

The expression for $\amp^{++}$ in Eq.~(\ref{amp++_final}) is finite for both the DVCS ($Q'^2\to 0$) and the TCS ($Q^2\to 0$) limits. In particular, for DVCS we have 
\begin{align}
    \amp^{++}_\textrm{DVCS} =&\ \lim_{\rho\to\xi}\amp^{++} \nonumber\\
    =&\ \int_{-1}^1dx\ \Bigg\{  -\left(1-\frac{t}{2\scale^2}\right)\frac{\hplus}{x-\xi+i0} \nonumber\\
    &\ - \frac{2t}{\scale^2}\Bigg[ \frac{1}{x+\xi}\Ln{\frac{x-\xi+i0}{-2\xi+i0}} +  \frac{L_{\rm DVCS}}{4}  \Bigg] \hplus \nonumber\\
    &\ + \frac{t}{\scale^2}\dxi\Bigg[ \Bigg(  \frac{\xi}{x+\xi}\Ln{\frac{x-\xi+i0}{-2\xi+i0}} + \frac{\xi L_{\rm DVCS}}{2}\Bigg)\hplus \Bigg] \nonumber\\
    &\ -\frac{\bp_\perp^2}{\scale^2}2\xi^3\dxi^2\Bigg[ \Bigg( 
    \frac{\xi}{x+\xi}\Ln{\frac{x-\xi+i0}{-2\xi+i0}} 
    + \frac{\xi L_{\rm DVCS}}{2}\Bigg)\hplus \Bigg]     \Bigg\} \nonumber\\
    &\ + O(\textrm{tw-6}) \,,\label{amp++DVCS_final}
\end{align}
where the scale of DVCS is simply $\scale^2 = Q^2+t$ and $L_\textrm{DVCS}$ is given by
\begin{align}
    L_\textrm{DVCS} & = \lim_{\rho\to\xi} L \nonumber\\
    & = \frac{4}{x-\xi}\left[ \Li{2}{\frac{x+\xi}{2\xi-i0}} - \Li{2}{1} \right]\,.\label{Ldvcs}
\end{align}
Equation~(\ref{amp++DVCS_final}) matches that of Ref.~\cite{Braun:2022qly} taking into account that the GPD used there ($H_{q\,\textrm{\cite{Braun:2022qly}}}$) relates to the one we employ here ($\hplus$) via: $\sum_q e_q^2 H_{q\,\textrm{\cite{Braun:2022qly}}} = \hplus/2$, where $e_q$ is the fractional electric charge of the quark of flavor~$q$.

In turn, the TCS limit of Eq.~(\ref{amp++_final}) corresponds to:
\begin{align}
    \amp^{++}_\textrm{TCS} =&\ \lim_{Q^2\to 0}\amp^{++} \nonumber\\
    =&\ \int_{-1}^1dx\ \Bigg\{ -\frac{1}{x+\xi(1 - 2t/\scale^2)+i0}\hplus + \frac{t}{2\scale^2}\frac{1}{x+\xi+i0}\hplus -\frac{2t}{\scale^2}\xi\dxi \Bigg(\frac{1}{x+\xi+i0}\hplus\Bigg)\nonumber\\
    &\ - \frac{2t}{\scale^2}\Bigg[ -\frac{1}{x-\xi}\Li{2}{-\frac{x-\xi}{2\xi+i0}}  -\frac{1}{x+\xi}\Bigg( \Li{2}{-\frac{x-\xi}{2\xi+i0}}-\Li{2}{1} \Bigg) + \frac{L_\textrm{TCS}}{4} \Bigg]\hplus \nonumber\\
    &\ +\frac{t}{\scale^2}\dxi\Bigg[ \Bigg( -\frac{2\xi}{x-\xi}\Li{2}{-\frac{x-\xi}{2\xi+i0}} - \frac{2\xi}{x+\xi}\Bigg( \Li{2}{-\frac{x-\xi}{2\xi+i0}}-\Li{2}{1} \Bigg) + \frac{\xi L_\textrm{TCS}}{2} \nonumber\\
    &\ \phantom{-\frac{t}{\scale^2}\dxi\Bigg[} +\frac{\xi}{x-\xi}\Ln{\frac{x+\xi+i0}{2\xi+i0}} \Bigg)\hplus \Bigg] \nonumber\\
    &\ -\frac{\bp_\perp^2}{\scale^2}2\xi^3\dxi^2\Bigg[ \Bigg( 
    -\frac{2\xi}{x-\xi}\Li{2}{-\frac{x-\xi}{2\xi+i0}} - \frac{2\xi}{x+\xi}\Bigg( 
    \Li{2}{-\frac{x-\xi}{2\xi+i0}}-\Li{2}{1} \Bigg)  + \frac{\xi L_\textrm{TCS}}{2}   \nonumber\\
    &\ \phantom{+\frac{\bp_\perp^2}{\scale^2}2\xi^3\dxi^2\Bigg[} + \frac{\xi}{x-\xi}\Ln{\frac{x+\xi+i0}{2\xi+i0}}  \Bigg)\hplus \Bigg] \Bigg\} + O(\textrm{tw-6})\,, \label{amp++TCS_final}
\end{align}
with the scale of TCS given by $\scale^2 = Q'^2+t$. The evaluation at $Q^2=0$ renders a value of $\rho$ that to the twist-4 accuracy can be approximated to $\rho \simeq -\xi(1 - 2t/\scale^2)$. 
The integral $L_\textrm{TCS}$ is given by taking the small incoming virtuality limit of Eq.~(\ref{Ldef}).

It is easy to see that at LT accuracy one recovers the usual result:
\begin{align}
    \left.\amp^{++}_\textrm{TCS}\right|_\textrm{LT} & = \lim_{Q^{\prime\,2}\to\infty} \amp^{++}_\textrm{TCS} =  -\int_{-1}^1dx\ \frac{\hplus}{x+\xi+i0} \,.
\end{align}

\subsection{Transverse-helicity flip amplitude, $\amp^{+-}$}\label{sec::A+-}

In this section we describe the transverse-helicity flip amplitude, denoted as $\amp^{+-}$. At LO, this amplitude appears as a kinematic higher-twist correction, whereas at NLO and LT accuracy it receives contributions from the gluon transversity GPD~\cite{Belitsky:2000jk}. Since we work at zeroth order in $\alpha_s$, the contribution detailed here will start at kinematic twist-4. 

Due to the form of the corresponding projector, $\Pi^{(+-)}_{\mu\nu}$ (\ref{Pi+-}), terms proportional to the metric, to longitudinal vectors or to an antisymmetric tensor  vanish. Consequently, at the twist-4 level, only the terms~$\sim z^\mu\partial^\nu\lop(\bu,0)$ and $\sim z^\nu\partial^\mu\lop(1,v)$ in the first line of the Compton tensor (\ref{ComptonCOPE_vectorPart}) contribute. After integration by parts, we get:
\begin{align}
     \left.\amp^{+-}\right|_\textrm{tw-4} & = \frac{1}{i\pi^2} i \int d^4z\ e^{iq'z}\, \Pi^{(+-)}_{\mu\nu} \frac{-4z^\mu z^\nu}{(-z^2+i0)^3} \left[ \int_0^1 du\ \lop (\bar u,0) + \int_0^1 dv\ \lop (1,v) \right] \nonumber\\
     & = -4\intba \phiplus\left( \int_0^1 \frac{du}{\bar{u}}\ \left.\frac{\Pi^{(+-)}_{\mu\nu}\ell ^\mu \ell ^\nu}{(q'-\ell)^2+i0}\right|_{\ell=\ell_{\bu,0}} + \int_0^1 \frac{dv}{\bar{v}}\ \left.\frac{\Pi^{(+-)}_{\mu\nu}\ell ^\mu \ell ^\nu}{(q'-\ell)^2+i0}\right|_{\ell=\ell_{1,v}} \right)\,.
\end{align}
Making use of the functional $\mathbb{I}_2$ in App.~\ref{app::DDs->GPDs}, we finally obtain:
\begin{align}
    \amp^{+-} =& -4\frac{\bp_\perp^2}{\scale^2}  (\xi^2\dxi)^2 \int_{-1}^{1} \frac{dx}{2\xi}\ \left( \wpbbiii - \wpbbi + 2\Ln{\frac{x-\rho+i0}{-2\xi}} \right) H^{(+)} + O(\textrm{tw-6})\,,
\end{align}
where functions $\wpbbi$ and $\wpbbiii$ are collected in Eqs.~(\ref{collectCoefficientsA++DDVCS}). The DVCS and TCS limits of $\amp^{+-}$ up to twist-4 are given by
\begin{align}
    \amp^{+-}_\textrm{DVCS} = -4\frac{\bp_\perp^2}{\scale^2}  (\xi^2\dxi)^2 \int_{-1}^{1} \frac{dx}{2\xi}\ \frac{2x}{x+\xi} \Ln{\frac{x-\xi+i0}{-2\xi}} H^{(+)} + O(\textrm{tw-6}) \,,\\
     \amp^{+-}_\textrm{TCS} = -4\frac{\bp_\perp^2}{\scale^2}  (\xi^2\dxi)^2 \int_{-1}^{1} \frac{dx}{2\xi}\ \frac{2x}{x-\xi} \Ln{\frac{x+\xi+i0}{2\xi}} H^{(+)} + O(\textrm{tw-6}) \,.
\end{align}
Our calculation of the DVCS amplitude above as a limiting case of DDVCS agrees with  the earlier publication~\cite{Braun:2022qly}. Note that the two limits are related by
\begin{equation}
\amp^{+-}_\textrm{TCS} = -\left(\amp^{+-}_\textrm{DVCS}\right)^*= \left.\left( \amp^{+-}_\textrm{DVCS} \right)\right|_{\substack{\xi\to -\xi\\Q^2\leftrightarrow -Q^{\prime\,2}}} \,,
\label{TransToTrans_DVCSvsTCS_relation}
\end{equation}
where $Q^2\leftrightarrow -Q'^2$ is a consequence of going from TCS (with a timelike virtuality) to DVCS (with a spacelike one), and $\xi\to -\xi$ is expected due to time reversal. This transformation would also change $\amp^{+-}$ to $\amp^{-+}$ but, thanks to parity conservation, they are equal. In Eq.~(\ref{TransToTrans_DVCSvsTCS_relation}), we took into account that $Q^2\leftrightarrow -Q'^2$ implies $\scale^2\to -\scale^2$ (up to a $t$ factor that we can ignore as it produces a higher twist term).

\subsection{Longitudinal-to-transverse and transverse-to-longitudinal  helicity flip amplitudes, $\amp^{0+}$ and $\amp^{+0}$}

For the cases of the helicity-flip amplitudes $\amp^{0+}$ and $\amp^{+0}$, the corresponding projectors satisfy
\begin{equation}
    \Pi^{(0+,\,+0)}_{\mu\nu}\left( g^{\mu\nu}, n^\mu n'^\nu, n^\nu n'^\mu \right) = 0\,,
\end{equation}
and the non-zero projections scale as
\begin{align}
    \Pi^{(0+)}_{\mu\nu}\left( q'^\nu\ell^\mu,\,\ell^\mu\ell^\nu,\,\Delta^\nu\ell^\mu \right) &\sim Q |\bp_\perp| \,,\\
    \Pi^{(+0)}_{\mu\nu}\left( q'^\mu\ell^\nu,\,\ell^\mu\ell^\nu,\,\Delta^\mu\ell^\nu \right) &\sim Q' |\bp_\perp| \,.
\end{align}
As a consequence, they generate twist-3 contributions whenever they encounter a factor that scales as $1/\scale^2$. In the context of the integrals $I_{n,m}$ (\ref{Inm}), $1/\scale^2$ factors appear through the terms $1/b$ and $1/(b+c)$. Therefore, to twist-3 accuracy:
\begin{align}\label{A0+/+0_tw3}
    \left.\amp^{0+,\,+0}\right|_\textrm{tw-3} \supset&\ -\frac{1}{i\pi^2}i\int d^4z\ e^{iq'z}\,\Pi_{\mu\nu}^{(0+,\,+0)} \nonumber\\
    &\ \times\Bigg\{ \frac{1}{(-z^2+i0)^2}\left[ 
    z^\mu\partial^\nu\int_0^1 du\ \lop(\bu, 0) + z^\nu(\partial^\mu - i\Delta^\mu)\int_0^1 dv\ \lop(1,v) \right] \nonumber\\
    &\ + \frac{i}{2(-z^2+i0)} (\Delta^\nu\partial^\mu - \Delta^\mu\partial^\nu)\int_0^1 du\int_0^{\bu} dv\ \lop(\bu,v) \Bigg\}\,.
\end{align}
Including the identity \mbox{$1=\int_{-1}^1 dx\ \deltaxba$} which relates DDs to GPDs, we obtain:
\begin{align}
    \amp^{+0} =& \int_{-1}^1dx\intba 4\phiplus\deltaxba\frac{i\beta Q'|\bp_\perp|}{\sqrt{2}\scale^2} \nonumber\\
    & \times\Bigg[\int_0^1 du\ \frac{\bu (x-\xi) - 2\rho}{\bu(x-\xi)+\xi-\rho+i0} + \int_0^1 dv\ \frac{\bv(x+\xi)}{\bv(x+\xi) - \xi-\rho+i0} \nonumber\\
    & \phantom{\times\Bigg[} -\int_0^1 du\int_0^\bu dv\ \frac{2\xi}{\bu(x-\xi)-v(x+\xi)+\xi-\rho+i0}\Bigg] + O(\textrm{tw-5})
\end{align}
and
\begin{align}
    \amp^{0+} =& \int_{-1}^1dx\intba 4\phiplus\deltaxba\frac{-\beta Q|\bp_\perp|}{\sqrt{2}\scale^2} \nonumber\\
    &\ \times \Bigg[ \int_0^1 du\ \frac{\bu(x-\xi)}{\bu(x-\xi)+\xi-\rho+i0} + \int_0^1dv\ \frac{\bv (x+\xi)-2\rho}{\bv(x+\xi)-\xi-\rho+i0} \nonumber\\
    &\ \phantom{\times\Bigg[}+ \int_0^1du\int_0^\bu dv\ \frac{2\xi}{\bu(x-\xi)-v(x+\xi)+\xi-\rho+i0} \Bigg] + O(\textrm{tw-5})\,.
\end{align}
Solving the integrals with respect to $u$ and $v$, and making use of the prescriptions described in App.~\ref{app::DDs->GPDs}:
\begin{align}
    \amp^{+0} & = \frac{-iQ'|\bp_\perp|}{\sqrt{2}\scale^2}\int_{-1}^1dx\ 4\xi^2\dxi\left( \frac{1}{x-\xi}\Ln{\frac{x-\rho+i0}{\xi-\rho+i0}}\hplus \right) + O(\textrm{tw-5})\,,\\
    \amp^{0+} & = \frac{-Q|\bp_\perp|}{\sqrt{2}\scale^2}\int_{-1}^1dx\  4\xi^2\dxi\left( \frac{1}{x+\xi}\Ln{\frac{x-\rho+i0}{-\xi-\rho+i0}}\hplus \right) + O(\textrm{tw-5})\,.\label{A0+_withDivergence}
\end{align}
Notice that $\amp^{+0}$ ($\amp^{0+})$ vanishes as we approach the DVCS (TCS) limit. This is expected since the outgoing (incoming) photon for DVCS (TCS) is real, hence it is transversely polarized.

However, as $\rho\to -\xi$ the amplitude $\amp^{0+}$ is logarithmically divergent. The point $\rho=-\xi$ is physical for a timelike-dominated DDVCS ($Q'^2 > Q^2$) as it only imposes $t=-Q^2$ and for the twist expansion to hold $Q^2<\scale^2 = Q'^2$. 

This divergence originates from the integral with operator $\lop(1,v)$ in Eq.~(\ref{A0+/+0_tw3}). The complete term producing the twist-3 contribution from $\lop(1,v)$, prior to the twist expansion, is:
\begin{align}\label{Idefinition}
    \inv =& \int_{-1}^1dx\intba 4\phiplus \deltaxba\frac{Q}{\sqrt{2}R|\bp_\perp|}\left.\int_0^1d\bv\ \frac{[(\ell q')-Q'^2](\ell\bp_\perp)}{\bv(a+b+c)}\right|_{\ell=\ell_{1,v}} \nonumber\\
    =& \int_{-1}^1dx\intba 4\phiplus \deltaxba\frac{1}{\sqrt{2}} \nonumber\\
    & \times\int_0^1d\bv\ \left[ \frac{\scale^2\rho}{2R\xi} - \bv\left( \frac{\beta}{2\xi}\left(1-\frac{\scale^2}{2R}\right) + \frac{\scale^2(x+\xi)}{4R\xi} \right) \right]\frac{\beta|\bp_\perp|Q}{\ell^2_{1,v} + \scale^2\mathbb{F}}\,,
\end{align}
where 
\begin{equation}
    \scale^2\mathbb{F} = 2R\left[ \frac{\bv\beta}{2\xi}\left(1-\frac{\scale^2}{2R}\right) + \bv\frac{\scale^2(x+\xi)}{4R\xi} - \frac{\scale^2(\xi+\rho-i0)}{4R\xi} \right]\,.
\end{equation}
Thus far, the approximation employed to solve this integral consisted of neglecting powers higher that $|t|/\scale^2$ and $M^2/\scale^2$ by removing $\ell_{1,v}^2$ from the denominator in Eq.~(\ref{Idefinition}). This approximation triggers the logarithmic divergence discussed above, which is a consequence of the non-commutativity of the integration in $\inv$ with respect to the auxiliary variable $\bv$ and the twist expansion to the accuracy kept so far (\mbox{twist-4}). Therefore, we need to add higher-order terms. Since $\bv\in(0,1)$ and
\begin{equation}\label{ell_1v}
    \ell_{1,v}^2 = t\left[ 1 - \frac{x+\xi-\beta}{\xi}\bv \right] + O(\bv^2 t,\, \bv^2\bp_\perp^2)\,,
\end{equation}
the next order in the approximation consists in keeping terms of the order of $|t|/\scale^2$ and $\bv|t|/\scale^2$, but dropping those proportional to $\bv^2|t|/\scale^2$ as they can be disregarded when compared to $\bv|t|/\scale^2$ for $\bv\in(0,1)$. 

Taking into account Eq.~(\ref{ell_1v}), we have that the denominator in $\inv$ is
\begin{equation}\label{ell_1v+Q2F}
    \ell_{1,v}^2+\scale^2\mathbb{F} = \scale^2\left\{ \frac{t}{\scale^2}\left[ 
    1-\frac{x+\xi}{\xi}\bv \right] + \frac{\bv\beta}{\xi}\frac{t}{\scale^2}\left[ 1-\frac{\xi-\rho}{2\xi} \right] + \bv\frac{x+\xi}{2\xi} - \frac{\xi+\rho-i0}{2\xi} + O\left( \bv^2 t/\scale^2,\,\bv^2\bp_\perp^2/\scale^2,\,\bv t^3/\scale^6 \right) \right\} \,.
\end{equation}
The divergence in Eq.~(\ref{A0+_withDivergence}) happened in the limit $\rho\to -\xi$. We observe that for this limit, the term proportional to $\bv\beta$ in the above equation vanishes, while for $\rho\neq -\xi$ it would constitute a term of twist higher than 4. Therefore, we can safely remove such term from Eq.~(\ref{ell_1v+Q2F}). With this new approach, $\inv$ takes the form:
\begin{align}\label{solvedI_nextOrderCorrection}
    \inv = & \int_{-1}^1dx\intba 4\phiplus \deltaxba\frac{1}{\sqrt{2}} \nonumber\\
    & \times\frac{-\beta|\bp_\perp|Q}{\scale^2}\left[ \frac{\xi-\rho}{x+\xi}\Ln{\frac{x(1-2t/\scale^2)-\rho+i0}{-\xi(1-2t/\scale^2)-\rho+i0}} +1 \right] + O(\textrm{tw-5})\,.
\end{align}
For the other contributions in Eq.~(\ref{A0+/+0_tw3}) which come from operators $\lop(\bu,0)$ and $\lop(\bu,v)$, the new approach renders the same results. The reason for this is that $\ell^2_{\bu,0}$ is directly proportional to $\bu^2$, while $\ell_{\bu,v}$ is proportional to $\bu^2$, $v^2$ and $\bu v$.

Introducing the expression (\ref{solvedI_nextOrderCorrection}) back in Eq.~(\ref{A0+/+0_tw3}) and using the mapping between DDs and GPDs given in App.~\ref{app::DDs->GPDs}, we get
\begin{align}
    \amp^{0+} =& \frac{-Q|\bp_\perp|}{\sqrt{2}\scale^2}\int_{-1}^1 dx\ 4\xi^2\dxi\Bigg( \frac{\hplus}{2\xi}\Bigg[ \frac{\xi+\rho}{x+\xi}\Ln{\frac{x-\rho+i0}{-\xi-\rho+i0}} \nonumber\\
    & + \frac{\xi-\rho}{x+\xi}\Ln{\frac{x(1-2t/\scale^2)-\rho+i0}{-\xi(1-2t/\scale^2)-\rho+i0}} \Bigg] \Bigg) + O(\textrm{tw-5})\,.
\end{align}
This new expression of $\amp^{0+}$ is regular in the whole phase-space of DDVCS. The logarithmic divergence happens now for 
\begin{equation}
    \rho \to -\xi(1-2t/\scale^2)\,,
\end{equation}
which is equivalent to $Q^2\to 0$ (TCS) and is regulated by the global factor $Q$ above. Indeed, we might as well write:
\begin{align}
    \amp^{0+} =& \frac{-Q|\bp_\perp|}{\sqrt{2}\scale^2}\int_{-1}^1 dx\ 4\xi^2\dxi\Bigg( \frac{\hplus}{2\xi}\Bigg[ \frac{\xi+\rho}{x+\xi}\Ln{\frac{x-\rho+i0}{-\xi-\rho+i0}} \nonumber\\
    & + \frac{\xi-\rho}{x+\xi}\Ln{\frac{x(1-2t/\scale^2)-\rho+i0}{-2\xi Q^2/\scale^2+i0}} \Bigg] \Bigg) + O(\textrm{tw-5})\,.
\end{align}
The TCS limit of this formula is trivially zero, while the DVCS one matches the result in Ref.~\cite{Braun:2022qly} and is given by:
\begin{equation}
    \amp^{0+}_\textrm{DVCS} = \frac{-Q|\bp_\perp|}{\sqrt{2}\scale^2}\int_{-1}^1 dx\ 4\xi^2\dxi\left( \frac{1}{x+\xi}\Ln{\frac{x-\xi+i0}{-2\xi+i0}}\hplus \right) + O(\textrm{tw-5})\,.
\end{equation}
In a similar manner, the DVCS limit of $\amp^{+0}$ is zero whereas
\begin{equation}
    \amp^{+0}_\textrm{TCS} = \frac{-iQ'|\bp_\perp|}{\sqrt{2}\scale^2}\int_{-1}^1dx\ 4\xi^2\dxi\left( \frac{1}{x-\xi}\Ln{\frac{x+\xi+i0}{2\xi+i0}}\hplus \right) + O(\textrm{tw-5})\,.
\end{equation}
Note that the two limits are related by 
\begin{equation}\label{LongToTrans_DVCSvsTCS_relation}
    \amp^{+0}_\textrm{TCS} = i\left(\amp^{0+}_\textrm{DVCS}\right)^*= \left.\left( \amp^{0+}_\textrm{DVCS} \right)\right|_{\substack{\xi\to -\xi\\Q\to iQ'}} \,.
\end{equation}

%% file: sec_numerics.tex
\section{Numerical estimates}
\label{sec:numerics}

\begin{figure}[!ht]
\centering
\includegraphics[width=0.9\textwidth]{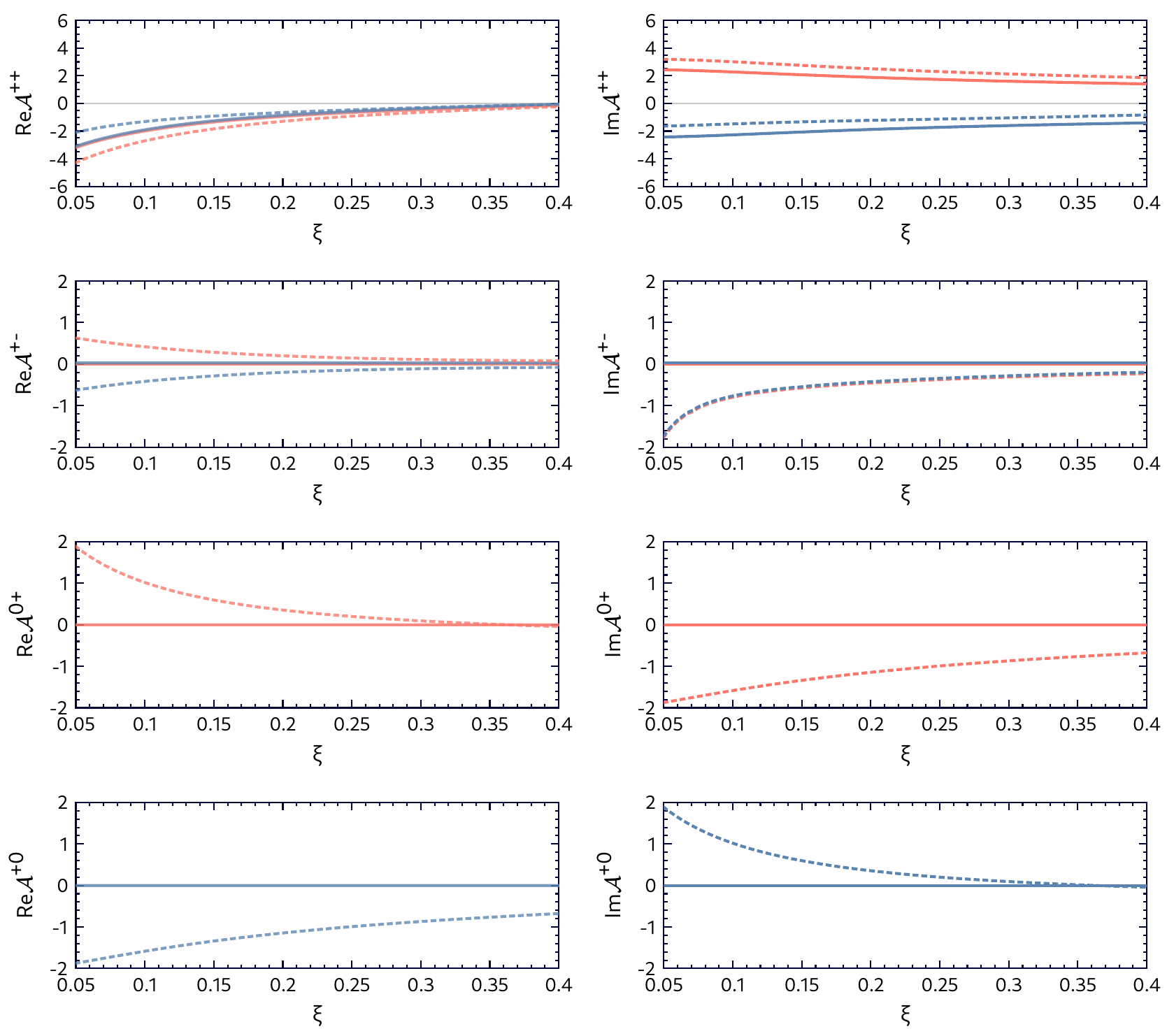}
\caption{Real and imaginary parts of $\amp^{++}$, $\amp^{+-}$, $\amp^{0+}$ and $\amp^{+0}$ amplitudes as a function of $\xi$ for $\scale^2 = 1.9\,\mathrm{GeV}^2$ and $t = -0.6\,\mathrm{GeV}^2$. Solid red (blue) curves correspond to DVCS (TCS) evaluated at twist-2 accuracy. The dotted counterparts represent evaluations that include corrections up to twist-4 accuracy.}
\label{fig:xi}
\end{figure}

\begin{figure}[!ht]
\centering
\includegraphics[width=0.9\textwidth]{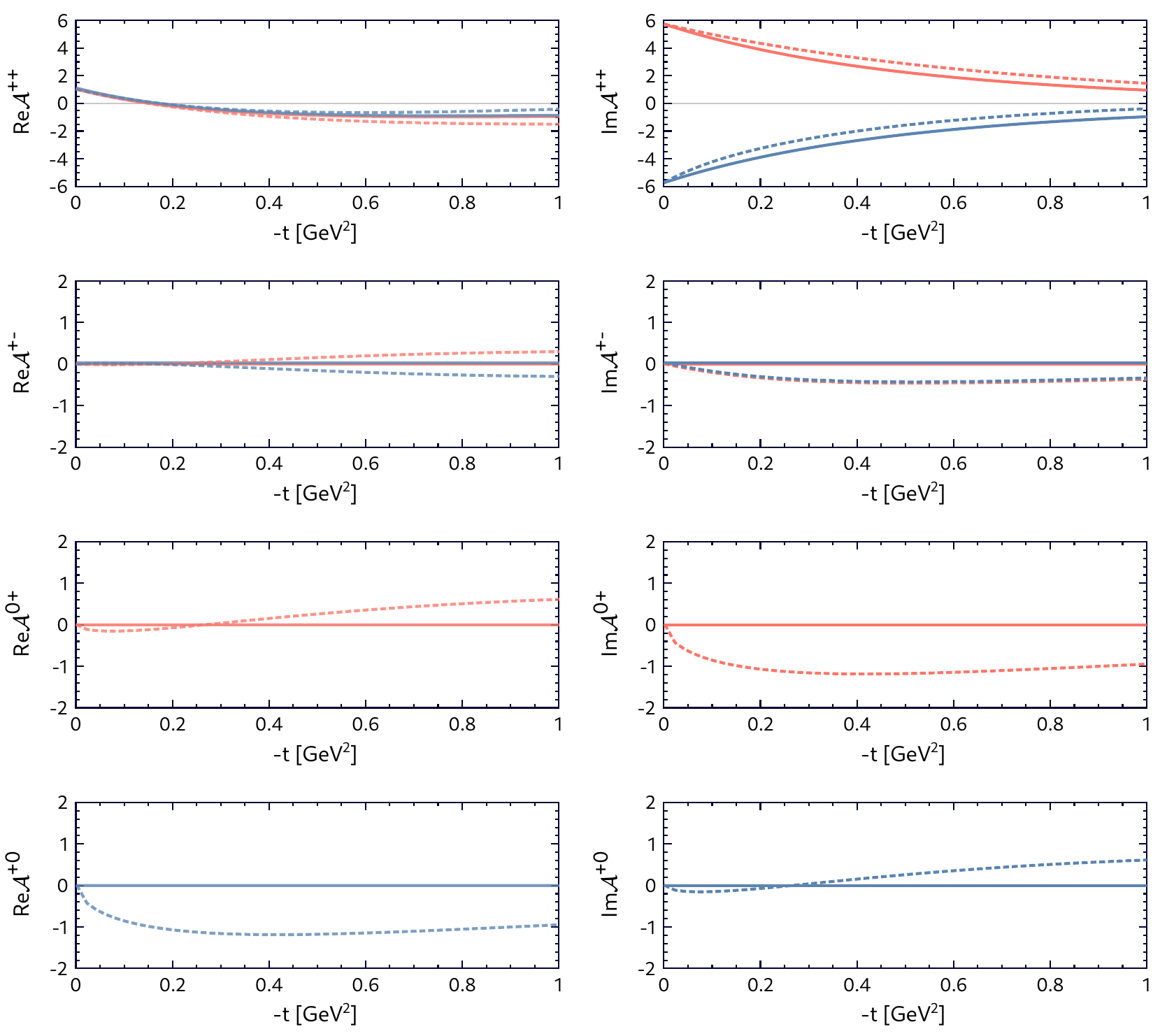}
\caption{The same as Fig.~\ref{fig:xi}, but as a function of $-t$ for $\scale^2 = 1.9\,\mathrm{GeV}^2$ and $\xi = 0.2$.}
\label{fig:mt}
\end{figure}

\begin{figure}[!ht]
\centering
\includegraphics[width=0.9\textwidth]{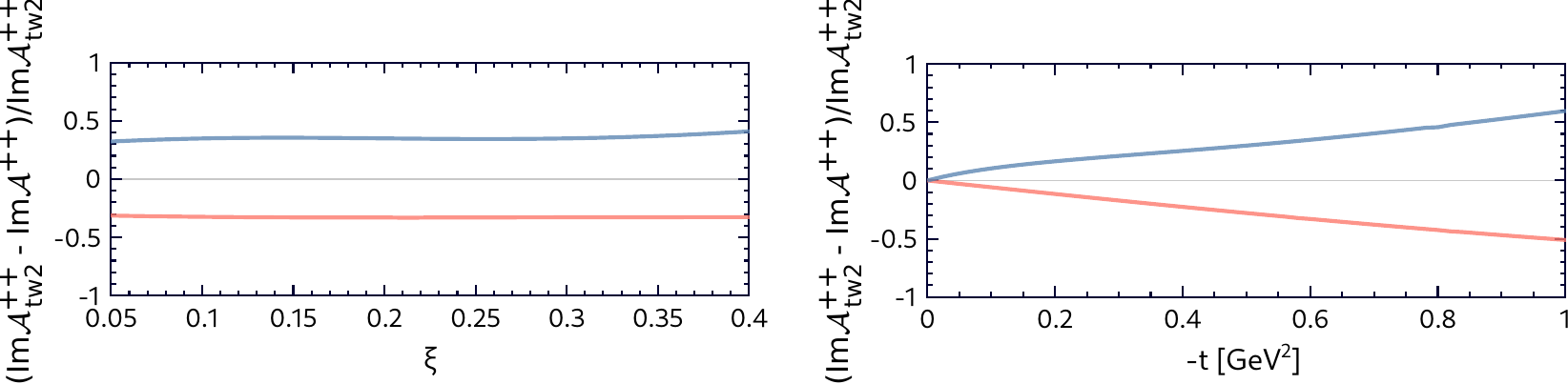}
\caption{The relative contribution of twist-4 corrections to twist-2 evaluation for the imaginary part of $\amp^{++}$ as a function of (left) $\xi$ at $t = -0.6\,\mathrm{GeV}^2$, and (right) $-t$ at $\xi=0.2$. Solid red (blue) curves correspond to DVCS (TCS). Both plots are for $\scale^2 = 1.9\,\mathrm{GeV}^2$.}
\label{fig:contrib}
\end{figure}

We now present estimates of the kinematic twist-3 and twist-4 contributions to the TCS helicity amplitudes on a pion target. We use the pion GPD model from Ref.~\cite{Chavez:2021llq}, which is based on Radyushkin's double distribution Ansatz~\cite{Musatov:1999xp}, the pion PDFs obtained in the xFitter framework~\cite{Novikov:2020snp}, and the dependence on the variable $t$ proposed in Ref.~\cite{Amrath:2008vx}. Our estimates are compared to the corresponding DVCS amplitudes. The results for a $^4\mathrm{He}$ target, along with their phenomenological implications, will be presented elsewhere. Figure~\ref{fig:xi} presents the $\xi$-dependence of the helicity amplitudes at $t = -0.6\,\mathrm{GeV}^2$, while in Fig.~\ref{fig:mt} we demonstrate the $t$-dependence at $\xi = 0.2$. Both figures are made for $\scale^2 = 1.9\,\mathrm{GeV}^2$. The corresponding plots showing the relative contribution of twist-4 effect for the dominant imaginary part of $\amp^{++}$ amplitude, i.e. $(\mathrm{Im}\amp^{++}|_{\mathrm{tw-2}}-\mathrm{Im}\amp^{++})/\mathrm{Im}\amp^{++}|_{\mathrm{tw-2}}$ ratio, is shown in Fig.~\ref{fig:contrib}. As expected, the effect does not depend significantly on  $\xi$, but it scales with $t$. For an exemplary point of $\xi = 0.2$ and $t = -0.6\,\mathrm{GeV}^2$ the inclusion of higher twist corrections leads to the change of the modulus of the amplitude by about $33\%$ for both DVCS and TCS. Note, however, that the kinematical twist-4 effects lead to a decrease in the modulus of the TCS amplitude, while they result in an increase in the DVCS amplitude. 

The presented estimates clearly demonstrate that kinematical higher-twist contributions are not negligible when pursuing a precise extraction of GPDs, particularly in tomography studies where the $t$-dependence is Fourier transformed into the impact distance $b_T$-dependence. 

Measuring the amplitudes that vanish at leading twist provides a new way to access leading-twist GPDs, however, these amplitudes are likely to also be influenced by genuine higher-twist effects that depend on higher-twist GPDs. The case of $\amp^{+-}$ is special and very interesting because, in addition to kinematic higher-twist contributions at LO involving quark chiral-even GPDs, it also includes leading-twist contributions at NLO (in $\alpha_s$) that are sensitive to gluon transversity GPDs.
For $\amp^{+-}$, as well as the pair of $\amp^{0+}$ and $\amp^{+0}$, the plots illustrate the symmetries between DVCS and TCS amplitudes reported in Eqs.~\eqref{TransToTrans_DVCSvsTCS_relation} and~\eqref{LongToTrans_DVCSvsTCS_relation}, respectively. For $\amp^{++}$, such relation is difficult to pinpoint, however, the inclusion of higher-twist effects clearly breaks the familiar LO LT relation $\amp^{++}_{\mathrm{DVCS}} = (\amp^{++}_{\mathrm{TCS}})^{*}$. We note that this relation is also broken after the inclusion of NLO corrections ~\cite{Mueller:2012sma,Grocholski:2019pqj}, making the study of the universality of GPDs overall more interesting than initially anticipated. 

\begin{figure}[!ht]
\centering
\includegraphics[width=0.9\textwidth]{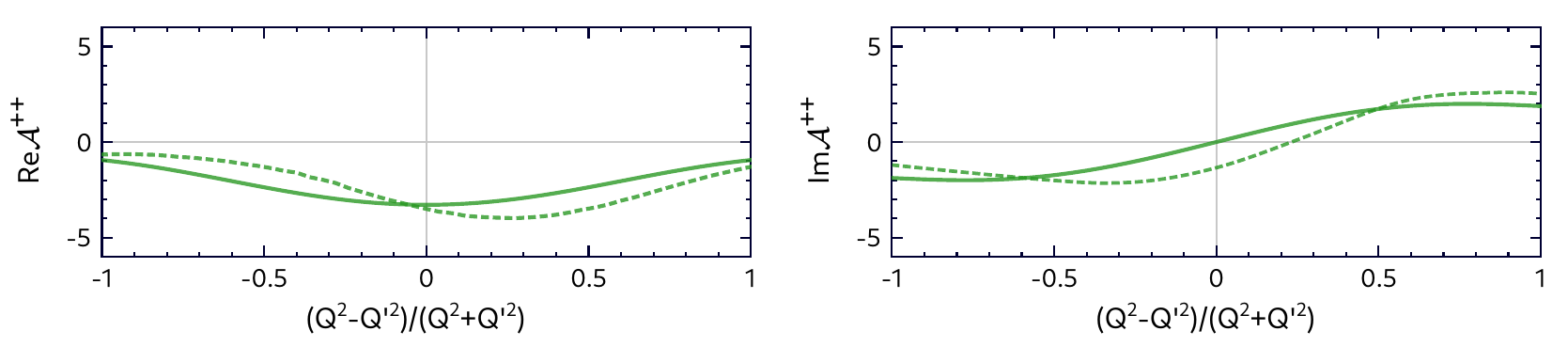}
\caption{Real and imaginary parts of $\amp^{++}$ amplitude as a function of $(Q^2-Q'^2)/(Q^2+Q'^2)$ ratio for $\scale^2 = 1.9\,\mathrm{GeV}^2$, $\xi=0.2$ and $t = -0.6\,\mathrm{GeV}^2$. Solid curves correspond to DDVCS evaluated at twist-2 accuracy. The dotted counterparts represent evaluations that include corrections up to twist-4 accuracy.}
\label{fig:ratio}
\end{figure}

While our work is dedicated to the TCS case and its relation to DVCS, it is important to emphasize that our formalism allows the study of a  transition between these two cases, for instance, by plotting the amplitudes as a function of $(Q^2-Q'^2)/(Q^2+Q'^2)$ ratio at fixed $\xi$, $t$ and $\scale^2$. For illustration, Fig.~\ref{fig:ratio} depicts the behavior of $\amp^{++}$ amplitude in this context. The limiting values $(Q^2-Q'^2)/(Q^2+Q'^2) = -1$ and $1$ correspond to the TCS and DVCS cases, respectively.

%% file: sec_conclusions.tex
\section{Conclusions}
\label{sec:conclusions}

By following the strategy based on the conformal operator-product expansion, we estimated the size of kinematic twist-3 and 4 corrections to the TCS process on a spin-0 target. Working within the general framework of two-photon scattering allowed us to compare our results to those for DVCS and to provide some insights into DDVCS as well. We found that the corrections are by no means negligible and they should therefore be taken into account in the extraction of GPDs from scattering cross-sections. On the other hand, the measurement of specific asymmetries, which vanish at leading twist but not at kinematic twist-3 and 4, such as those originating from $\amp^{+-}$ and $\amp^{0+}$ helicity amplitudes, provides a new way to access the same GPDs.

In this work, we did not attempt to write the corresponding amplitudes for a nucleon target. However, most of the analytical work presented here will be useful for that case. Extending the study to a spin-1 target, such as the deuteron, would also be valuable since deuteron GPDs have very interesting new facets \cite{Berger:2001zb,Cano:2003ju}, and the DVCS and TCS processes on the deuteron should soon be experimentally accessible~\cite{CLAS:2024qhy}.

We note that many questions related to the line of research we are pursuing here remain open. On the theory side, since NLO corrections have been shown to be quite sizable \cite{Moutarde:2013qs}, the extension of our calculation of the kinematical corrections to an amplitude calculated at NLO is obviously needed. Although corrections to the Born approximation render QCD non-conformal, the conformal techniques can still be useful by considering the fixed points of the $\beta$-function and the renormalization group equations~\cite{Braun:2016qlg}. In addition, a reliable extraction of GPDs from data will definitely require trustable estimates of genuine higher twist corrections to the leading-twist analysis, which may be aided by lattice computations of higher twist GPDs (see for instance Ref.~\cite{Bhattacharya:2023nmv}). On the experimental side, one should carefully study the phenomenological implications of higher twist effects for DVCS and TCS observables. For spin-0 targets, one should not overlook the $^4\mathrm{He}$ nucleus, as related experimental data are already available for DVCS~\cite{CLAS:2021ovm}. We hope that some of these open questions will be resolved before the next generation of experiments~\cite{AbdulKhalek:2021gbh,Anderle:2021wcy,Accardi:2023chb} provides a rich harvest of data on exclusive reactions.

%% file: acknowledgements.tex
\paragraph*{Acknowledgements.}

We   thank V.~M.~Braun and A.~N.~Manashov for fruitful discussions and for the invitation of one of us (V.M.F.) to the University of Regensburg, which helped us to clarify certain aspects of the application of the conformal methods in QCD. We also acknowledge very useful discussions with Q.T.~Song, L.~Szymanowski, C.~Mezrag and J-M.~Morgado Chavez. This research was funded, in part, by l'Agence Nationale de la Recherche (ANR), project ANR-23-CE31-0019. For the purpose of open access, the authors have applied a CC-BY copyright licence to any Author Accepted Manuscript (AAM) version arising from this submission. The work of P.~S. and J.~W.
was supported by the Grant No.~2024/53/B/ST2/00968
of the National Science Centre.

%% file: appendices/AppendixDD.tex
\section{ DDs and GPDs}\label{app::DDs}

According to \cite{Polyakov:1999gs}, one may consider the following double distribution (DD) representation for the GPDs of a  (pseudo-)scalar\footnote{Processes such as DVCS, TCS and DDVCS project the C-even part of GPDs (this is $H^{(+)}_f$).} 
hadron:
\begin{equation}\label{corrOp_f_fgSeparated}
    \langle p'| \Op_f(\lambda_1, \lambda_2) |p\rangle = \intba e^{-i\ell_{\lambda_1,\lambda_2}n}\left[ 2(\bp n)h_f(\beta,\alpha,t) - (\Delta n) g_f(\beta,\alpha,t) \right]\,,
\end{equation}
where $h_f, g_f$ are the so-called double distributions for quark flavor $f$, the vector $\ell_{\lambda_1,\lambda_2}$ is
\begin{equation}
    \ell_{\lambda_1,\lambda_2} = -\lambda_1\Delta - \lambda_{12}\left[ 
\beta\bp - \frac{1}{2}(\alpha+1)\Delta \right]\,,
\end{equation}
and the domain $\mathbb{D}$ for integration with respect to variables $\alpha$ and $\beta$ is defined as 
\begin{equation}
    \intba = \int_{-1}^1 d\beta\int_{|\beta|-1}^{1-|\beta|} d\alpha\,.
\end{equation}
The function $g_f$ is related to the $D$-term~\cite{Polyakov:1999gs}. Taking the Fourier transform to isolate the GPD, one finds
\begin{equation}\label{H+_via_h&g}
    \frac{1}{2}H^{(+)}_f(x,\xi,t) = \intba \delta(x-\beta-\alpha\xi)\left[ h_f(\beta,\alpha,t) + \xi g_f(\beta,\alpha,t)\right]\,.
\end{equation}
Nevertheless, Eq.~(\ref{corrOp_f_fgSeparated}) can be further simplified if one assumes $h_f, g_f$ to vanish in the boundaries of the domain $\mathbb{D}$. Expressing $\bp n$ and $\Delta n$ by derivatives with respect to $\beta$ and $\alpha$, respectively, acting on the exponential of Eq.~(\ref{corrOp_f_fgSeparated}), one gets via integration by parts:
\begin{equation}\label{corrOp_f}
    \langle p'| \Op_f(\lambda_1, \lambda_2) |p\rangle = \frac{2i}{\lambda_{12}}\intba e^{-i\ell_{\lambda_1,\lambda_2}n}\Phi_f^{(+)}(\beta,\alpha,t)\,,\quad z_i=\lambda_i n \,,
\end{equation}
where
\begin{equation}\label{PhiPlusFdef}
    \Phi_f^{(+)}(\beta,\alpha,t) = \partial_\beta h_f + \partial_\alpha g_f\,.
\end{equation}
The function $\phiplus$ is even under $(\beta,\alpha)\rightarrow (-\beta,-\alpha)$:
\begin{equation}
    \Phi_f^{(+)}(-\beta,-\alpha,t) = \Phi_f^{(+)}(\beta,\alpha,t)
\end{equation}
and consequently,
\begin{equation}\label{intbaAlphaPhi=0_and_intbaBetaPhi=0}
    \intba\alpha\Phi^{(+)}_f(\beta,\alpha,t) = 0\,,\quad \intba\beta\Phi^{(+)}_f(\beta,\alpha,t) = 0\,.
\end{equation}
Because $h_f$ and $g_f$ vanish in the frontier of $\mathbb{D}$, it follows:
\begin{equation}\label{intbaPhi=0}
    \intba \Phi^{(+)}_f(\beta,\alpha,t) = 0\,.
\end{equation}

When summing over flavors, we will use the notation:
\begin{equation}\label{PhiPlus}
    \PhiPlus = \sum_f \left(\frac{e_f}{e}\right)^2\PhiPlusF
\end{equation}
and 
\begin{equation}
    \HplusX = \sum_f \left(\frac{e_f}{e}\right)^2\HplusFx\,.
\end{equation}
The function $\phiplus$ can be related to the GPD $\hplus$ by introducing the identity
\begin{equation}\label{1=intDeltaBAXi}
    1=\int_{-1}^1 dx\ \delta(x-\beta-\alpha\xi)
\end{equation}
in Eq.~(\ref{corrOp_f}):
\begin{equation}\label{dH+/dx}
    \frac{1}{2}\partial_x H^{(+)}(x,\xi,t) = \intba \delta(x-\beta-\alpha\xi)\Phi^{(+)}(\beta,\alpha,t)\,.
\end{equation}

%% file: appendices/DDsToGPDs.tex
\section{Prescription to map DDs to GPDs in convolutions}\label{app::DDs->GPDs}

Up to kinematic twist-4, the hard-coefficient functions are given by the integrals
\begin{align}
    \mathbb{I}_0[g] & = \int_{-1}^1dx\intba 2\phiplus\deltaxba g(x,\xi,\rho)\,, \\
    \mathbb{I}_k[f] & = \int_{-1}^1dx\intba 2\phiplus\deltaxba \frac{(\xi\beta)^k}{\xi^2} f(x/\xi,\rho/\xi)\,,\quad k>0
\end{align}
For our purposes, it is enough to solve $\mathbb{I}_0$, $\mathbb{I}_1$ and $\mathbb{I}_2$.
Functional $\mathbb{I}_0$ makes use of relation (\ref{dH+/dx}) and integration by parts. The result, for any function $g=g(x,\xi,\rho)$, is:
\begin{equation}
    \mathbb{I}_0[g] = -\int_{-1}^1dx\ \left(\dx g \right)\hplus  \,.
\end{equation}
For $k\in\{1,2\}$, we have the following solutions:
\begin{align}
    \mathbb{I}_1[f] & = \int_{-1}^1dx\ \left[ \dxi\left(f\hplus\right) - \frac{f}{\xi}\hplus \right]\,, \label{prescription_I1} \\
    \mathbb{I}_2[f] & = \xi^3\dxi^2\int_{-1}^1dx\ \frac{\ycal[f]\hplus}{\xi}\,, \label{prescription_I2}
\end{align}
where $f = f(x/\xi,\rho/\xi)$ and 
\begin{equation}\label{ycal_def}
    \ycal[f] = \int_{x}^1dx'\ f(x'/\xi,\rho/\xi)\,.
\end{equation}

For $k=1$, note that
\begin{equation}
    \beta\deltaxba = (x\dx+\xi\dxi)\thetaxba\,,
\end{equation}
which together with 
\begin{equation}\label{thetaToGPD}
    \intba \thetaxba\PhiPlus = \frac{1}{2}\HplusX\,,
\end{equation}
allows to write
\begin{equation}
    \mathbb{I}_1[f] = \int_{-1}^1dx\ \frac{f}{\xi}(x\dx+\xi\dxi)\HplusX \,.
\end{equation}
Integration by parts plus the relation
\begin{equation}\label{dxiToDx}
    \dxi f(x/\xi,\rho/\xi) = -\frac{x}{\xi}\dx f(x/\xi,\rho/\xi)\,,
\end{equation}
which can be used to trade $\dxi$ by $\dx$ in order to solve $\int dx$, renders
\begin{equation}
    \mathbb{I}_1[f] = \int_{-1}^1dx\ \left[ \dxi\left(f\hplus\right) - \frac{f}{\xi}\hplus \right]\,.
\end{equation}

For $k=2$, note that
\begin{equation}
    \beta^2\deltaxba = (x^2\dx+2x\xi\dxi)\thetaxba + \xi^2\dxi^2\int_{-1}^xdx'\ \theta(x'-\beta-\alpha\xi)\,,
\end{equation}
and that we can exchange the order of integration over $x$ and $x'$ variables via
\begin{align}
    \int_{-1}^1dx\int_{-1}^x dx'\ C(x)\hplus(x',\xi,t) &=  \int_{-1}^1dx'\int_{x'}^1 dx\ C(x)\hplus(x',\xi,t) \nonumber\\
    &= \int_{-1}^1dx\ \hplus(x,\xi,t)\int_{x}^1dx'\ C(x')\,,
\end{align}
for any function $C(x)$. After integration by parts, these formulas together with Eqs.~(\ref{thetaToGPD}) and (\ref{dxiToDx}) yield:
\begin{align}
    \mathbb{I}_2[f] & = \xi^3\dxi^2\int_{-1}^1dx\ \frac{\HplusX}{\xi}\int_{x}^1dx'\ f(x'/\xi,\rho/\xi)\\
    & = \xi^3\dxi^2\int_{-1}^1dx\ \frac{\ycal[f]\hplus}{\xi}\,,
\end{align}
where we identified $\ycal[f]$ as in Eq.~(\ref{ycal_def}).

%% file: appendices/A00.tex
\section{Longitudinal-helicity conserving amplitude, $\amp^{00}$}\label{Appendix-A00}

Although the $\amp^{00}$ amplitude does not exist in either the TCS or DVCS limit, we write here for completeness its twist-$4$ contribution to the DDVCS scattering amplitude.
The projector onto $\amp^{00}$, vid.~Eq.~(\ref{Pi00}), is an antisymmetric tensor built out of the longitudinal vectors $q$ and $q'$. Hence, when contracted with the projector, transverse momenta and symmetric tensors vanish. Therefore, the longitudinal-helicity conserving amplitude takes the form:
\begin{align}\label{A00}
    \amp^{00} =& \Pi^{(00)}_{\mu\nu}T^{\mu\nu} = \Pi^{(00)}_{\mu\nu}\frac{1}{i\pi^2}i\int d^4z\ e^{iq'z}  \nonumber\\
    & \times\Bigg\{\frac{-1}{(-z^2+i0)^2}\left[ z^\mu\partial^\nu\int_0^1du\ \lop(\bu,0) + z^\nu(\partial^\mu-i\Delta^\mu)\int_0^1 dv\ \lop(1,v) \right] \nonumber\\
    & - \frac{1}{-z^2+i0}\left[ i\Delta^\nu\partial^\mu\int_0^1du\int_0^\bu dv\ \lop(\bu,v) - \frac{t}{4}z^\mu\partial^\nu\int_0^1du\ u\int_0^\bu dv\ \lop(\bu,v) \right] \nonumber\\
    & -\frac{iz^\mu\Delta^\nu}{4(-z^2+i0)}\int_0^1du\int_0^\bu dv\ \left[C_{(1),\,z\Delta}(\bu,v)\lop_1(\bu,v) + C_{(2),\,z\Delta}(\bu,v)\lop_2(\bu,v) \right] \nonumber\\
    & + \frac{z^\nu\partial^\mu}{-z^2+i0}\int_0^1 du\int_0^\bu dv\ \left[ C_{(1),\,z\partial}(\bu,v)\lop_1(\bu,v) + C_{(2),\,z\partial}(\bu,v)\lop_2(\bu,v) \right]  \Bigg\} \,.
\end{align}
where we defined the following conformal weights:
\begin{align}\label{C_(k),tensor}
    C_{(1),\,z\Delta}(\bu,v) &= \Ln{\frac{\bu-v}{\bu(1-v)}} + 2\frac{(1-\bu)v}{\bu-v} + 2\frac{v}{1-v}\left(1+\frac{2(1-\bu)v}{\bu-v}\right)\,, \nonumber\\
    C_{(2),\,z\Delta}(\bu,v) &= \frac{v}{1-v} - 2\frac{\bu v}{\bu-v} - \left(\frac{v}{1-v}\right)^2 - \frac{v}{1-v}\delta(1-\bu)\,,\nonumber\\
    C_{(1),\,z\partial}(\bu,v) &= -\frac{(1-\bu)v}{\bu-v}-\frac{1}{2}\left[ \left(1+\frac{2(1-\bu)v}{\bu-v}\right)\frac{v}{1-v} + \ln\bu + 1 - \bu \right]\,,\nonumber\\
    C_{(2),\,z\partial}(\bu,v) &= \frac{v}{\bu-v}+\frac{1}{4}\left(\frac{v}{1-v}\right)^2+\frac{1}{4}\frac{v}{1-v}\delta(1-\bu)-\frac{1}{2}\bu + \frac{1}{4}\,.
\end{align}
Amplitude $\amp^{00}$ may be decomposed as:
\begin{align}
    \amp^{00} =&\ \amp^{00}_{(0),\,z\partial} + \amp^{00}_{(0),\,z(\partial-i\Delta)} + \amp^{00}_{(0),\,\Delta\partial} + \amp^{00}_{(0),\,t z\partial} + \amp^{00}_{(1),\,z\Delta} + \amp^{00}_{(2),\,z\Delta} + \amp^{00}_{(1),\,z\partial} + \amp^{00}_{(2),\,z\partial}\,,
\end{align}
where after Fourier transforms, power expansion up to twist-4 and the use of the functionals described in App.~\ref{app::DDs->GPDs}, the terms on the RHS are expressed as:
\begin{align}\label{A00_0zder}
     &\amp^{00}_{(0),\,z\partial} = 2\frac{iQQ'}{\scale^2}\int_{-1}^1dx\int_0^1 du\ \frac{\xi-\rho}{2\xi} && \nonumber\\
     &\qquad \times\Bigg\{ \frac{t}{\scale^2}\Bigg[ 3N(\bu,0)-4(\xi-\rho)(N(\bu,0))^2+2(\xi-\rho)^2(N(\bu,0))^3 - 4\widetilde{N}(\bu,0) \Bigg]\hplus && \nonumber\\
     &\qquad -\frac{t}{\scale^2}\dxi\left( \xi\left[ 2N(\bu,0)-4\widetilde{N}(\bu,0) \right]\hplus \right) \nonumber\\
     &\qquad  +\frac{\bp_\perp^2}{\scale^2}2\xi^3\dxi^2\left( \xi\left[ 2N(\bu,0) - 4 \widetilde{N}(\bu,0) \right]\hplus \right)  \Bigg\} + O(\textrm{tw-6})\,, &&
\end{align}
\begin{align}
    & \amp^{00}_{(0),\,z(\partial-i\Delta)} = 2\frac{iQQ'}{\scale^2} \int_{-1}^1dx\int_0^1dv\ &&\nonumber\\
    & \qquad \times\Bigg\{ \frac{t}{\scale^2}\Bigg[ \frac{\xi+2\rho}{\xi}N(1,v) + (\xi-\rho)\left(\frac{2(\xi-\rho)}{\xi}-3\right)(N(1,v))^2 && \nonumber\\
    & \qquad\qquad + \frac{(\xi-\rho)^2(\xi+\rho)}{\xi}(N(1,v))^3 - \frac{2\rho}{\xi} \widetilde{N}(1,v) \Bigg]\hplus &&\nonumber\\
    &\qquad -\frac{t}{\scale^2}\dxi\left( \left[ (\xi+\rho)N(1,v) - 2\rho\widetilde{N}(1,v) \right]\hplus \right) &&\nonumber\\
    &\qquad + \frac{\bp_\perp^2}{\scale^2}2\xi^3\dxi^2\left( \left[ (\xi+\rho)N(1,v) - 2\rho\widetilde{N}(1,v) \right]\hplus \right) \Bigg\} + O(\textrm{tw-6})\,,  &&
\end{align}
\begin{align}
    & \amp^{00}_{(0),\,\Delta\partial} = 2\frac{iQQ'}{\scale^2}\int_{-1}^1dx\int_0^1du\int_0^\bu dv && \nonumber\\
    &\qquad \times\Bigg\{ \frac{-2t}{\scale^2} \left[ N(\bu,v) - (\xi-\rho)(N(\bu,v))^2 - 2\widetilde{N}(\bu,v) \right] \hplus -\frac{t}{\scale^2}\dxi\left( 4\xi\widetilde{N}(\bu,v)\hplus \right) &&\nonumber\\
    &\qquad +\frac{\bp_\perp^2}{\scale^2}2\xi^3\dxi^2\left( 4\xi\widetilde{N}(\bu,v)\hplus \right) \Bigg\} + O(\textrm{tw-6})\,,
\end{align}
\begin{align}
    & \amp^{00}_{(0),\,tz\partial} = 2\frac{iQQ'}{\scale^2}\int_{-1}^1dx\int_0^1du\int_0^\bu dv\ u\frac{2\xi t}{\scale^2}\left[ (N(\bu,v))^2 - 2(\xi-\rho)(N(\bu,v))^3 \right]\hplus &&\nonumber\\
    &\qquad + O(\textrm{tw-6})\,,   &&
\end{align}
\begin{align}\label{A00_1zDelta}
    & \amp^{00}_{(1),\,z\Delta} = 2\frac{iQQ'}{\scale^2}\int_{-1}^1dx\int_0^1du\int_0^\bu dv\ C_{(1),\,z\Delta}(\bu,v) && \nonumber\\
    &\qquad \times\Bigg\{ \frac{t}{\scale^2}\left[ \frac{2\xi}{\xi-\rho+i0}N(\bu,v) - 2\xi(\xi-\rho)(N(\bu,v))^3 \right]\hplus && \nonumber\\
    &\qquad -\frac{t}{\scale^2}\frac{2\xi}{\xi-\rho+i0}\dxi\left( \xi N(\bu,v)\hplus \right) + \frac{\bp_\perp^2}{\scale^2}\frac{2\xi}{\xi-\rho+i0}2\xi^3\dxi^2\left( \xi N(\bu,v)\hplus \right) \Bigg\} && \nonumber\\
    &\qquad+ O(\textrm{tw-6}) \,, &&
\end{align}   
\begin{align}\label{A00_2zDelta}
    & \amp^{00}_{(2),\,z\Delta} = 2\frac{iQQ'}{\scale^2}\int_{-1}^1dx\int_0^1du\int_0^\bu dv\ C_{(2),\,z\Delta}(\bu,v) && \nonumber\\
    &\qquad \times\Bigg\{ \frac{t}{\scale^2}\left[ \frac{2\xi}{\xi-\rho+i0}N(\bu,v) + 2\xi(\xi+\rho)(N(\bu,v))^3 \right]\hplus && \nonumber\\
    &\qquad -\frac{t}{\scale^2}\frac{2\xi}{\xi-\rho+i0}\dxi\left( \xi N(\bu,v)\hplus \right) + \frac{\bp_\perp^2}{\scale^2}\frac{2\xi}{\xi-\rho+i0}2\xi^3\dxi^2\left( \xi N(\bu,v)\hplus \right) \Bigg\} && \nonumber\\
    &\qquad+ O(\textrm{tw-6}) \,, &&
\end{align}   
\begin{align}
    &\amp^{00}_{(1),\,z\partial} = 2\frac{iQQ'}{\scale^2}\int_{-1}^1dx\int_0^1du\int_0^\bu dv\ C_{(1),\,z\partial}(\bu,v)   && \nonumber\\
    & \qquad \times\Bigg\{ \frac{2t}{\scale^2} \left[ 3N(\bu,v) + (\xi-\rho)(N(\bu,v))^2 -2(\xi-\rho)^2(N(\bu,v))^3 - 2\widetilde{N}(\bu,v) \right]\hplus &&\nonumber\\
    &\qquad -\frac{t}{\scale^2}\dxi\left( 4\xi\left[ N(\bu,v) - \widetilde{N}(\bu,v) \right]\hplus \right) + \frac{\bp_\perp^2}{\scale^2}2\xi^3\dxi^2\left( 4\xi\left[ N(\bu,v) - \widetilde{N}(\bu,v) \right]\hplus \right)  \Bigg\} && \nonumber\\
    &\qquad + O(\textrm{tw-6})\,,&&
\end{align}
\begin{align}\label{A00_2zder}
    &\amp^{00}_{(2),\,z\partial} = 2\frac{iQQ'}{\scale^2}\int_{-1}^1dx\int_0^1du\int_0^\bu dv\ C_{(2),\,z\partial}(\bu,v)   && \nonumber\\
    & \qquad \times\Bigg\{ \frac{2t}{\scale^2} \left[ 3N(\bu,v) - (\xi+\rho)(N(\bu,v))^2 +2(\xi-\rho)(\xi+\rho)(N(\bu,v))^3 - 2\widetilde{N}(\bu,v) \right]\hplus &&\nonumber\\
    &\qquad -\frac{t}{\scale^2}\dxi\left( 4\xi\left[ N(\bu,v) - \widetilde{N}(\bu,v) \right]\hplus \right) + \frac{\bp_\perp^2}{\scale^2}2\xi^3\dxi^2\left( 4\xi\left[ N(\bu,v) - \widetilde{N}(\bu,v) \right]\hplus \right)  \Bigg\} && \nonumber\\
    &\qquad + O(\textrm{tw-6})\,.&&
\end{align}
Note that the hard coefficients have been simplified to integrals with respect to $u$ and $v$ of two functions
\begin{align}\label{N_widetildeN}
    N(\lambda_1,\lambda_2) & = \frac{1}{\lambda_1(x-\xi)-\lambda_2(x+\xi)+\xi-\rho+i0}\,, \nonumber \\
    \widetilde{N}(\lambda_1,\lambda_2) & = \frac{1}{\lambda_1(x-\xi)-\lambda_2(x+\xi)}\Ln{1+\frac{\lambda_1(x-\xi)-\lambda_2(x+\xi)}{\xi-\rho+i0}}\,,
\end{align}
weighted with the conformal factors (\ref{C_(k),tensor}). The complexity of the functions~(\ref{C_(k),tensor}) and (\ref{N_widetildeN}) render long expressions for the hard coefficients of $\amp^{00}$ that are difficult to manage. Consequently, it seems more practical to express the subamplitudes of $\amp^{00}$ by means of hard coefficients given through integrals over $u$ and $v$ with the conformal weights (\ref{C_(k),tensor}), as illustrated in Eqs.~(\ref{A00_0zder}) to (\ref{A00_2zder}).

%% file: main.bbl
\begin{thebibliography}{58}
\expandafter\ifx\csname natexlab\endcsname\relax\def\natexlab#1{#1}\fi
\expandafter\ifx\csname bibnamefont\endcsname\relax
  \def\bibnamefont#1{#1}\fi
\expandafter\ifx\csname bibfnamefont\endcsname\relax
  \def\bibfnamefont#1{#1}\fi
\expandafter\ifx\csname citenamefont\endcsname\relax
  \def\citenamefont#1{#1}\fi
\expandafter\ifx\csname url\endcsname\relax
  \def\url#1{\texttt{#1}}\fi
\expandafter\ifx\csname urlprefix\endcsname\relax\def\urlprefix{URL }\fi
\providecommand{\bibinfo}[2]{#2}
\providecommand{\eprint}[2][]{\url{#2}}

\bibitem[{\citenamefont{M\"uller et~al.}(1994)\citenamefont{M\"uller, Robaschik, Geyer, Dittes, and Ho\v{r}ej\v{s}i}}]{Muller:1994ses}
\bibinfo{author}{\bibfnamefont{D.}~\bibnamefont{M\"uller}}, \bibinfo{author}{\bibfnamefont{D.}~\bibnamefont{Robaschik}}, \bibinfo{author}{\bibfnamefont{B.}~\bibnamefont{Geyer}}, \bibinfo{author}{\bibfnamefont{F.~M.} \bibnamefont{Dittes}}, \bibnamefont{and} \bibinfo{author}{\bibfnamefont{J.}~\bibnamefont{Ho\v{r}ej\v{s}i}}, \bibinfo{journal}{Fortsch. Phys.} \textbf{\bibinfo{volume}{42}}, \bibinfo{pages}{101} (\bibinfo{year}{1994}), \eprint{hep-ph/9812448}.

\bibitem[{\citenamefont{Ji}(1997)}]{Ji:1996nm}
\bibinfo{author}{\bibfnamefont{X.-D.} \bibnamefont{Ji}}, \bibinfo{journal}{Phys. Rev. D} \textbf{\bibinfo{volume}{55}}, \bibinfo{pages}{7114} (\bibinfo{year}{1997}), \eprint{9609381}.

\bibitem[{\citenamefont{Diehl}(2002)}]{Diehl:2002he}
\bibinfo{author}{\bibfnamefont{M.}~\bibnamefont{Diehl}}, \bibinfo{journal}{Eur. Phys. J. C} \textbf{\bibinfo{volume}{25}}, \bibinfo{pages}{223} (\bibinfo{year}{2002}), \bibinfo{note}{[Erratum: Eur.Phys.J.C 31, 277--278 (2003)]}, \eprint{hep-ph/0205208}.

\bibitem[{\citenamefont{Belitsky and Radyushkin}(2005)}]{Belitsky:2005qn}
\bibinfo{author}{\bibfnamefont{A.~V.} \bibnamefont{Belitsky}} \bibnamefont{and} \bibinfo{author}{\bibfnamefont{A.~V.} \bibnamefont{Radyushkin}}, \bibinfo{journal}{Phys. Rept.} \textbf{\bibinfo{volume}{418}}, \bibinfo{pages}{1} (\bibinfo{year}{2005}), \eprint{hep-ph/0504030}.

\bibitem[{\citenamefont{Burkardt}(2003)}]{Burkardt:2002hr}
\bibinfo{author}{\bibfnamefont{M.}~\bibnamefont{Burkardt}}, \bibinfo{journal}{Int. J. Mod. Phys. A} \textbf{\bibinfo{volume}{18}}, \bibinfo{pages}{173} (\bibinfo{year}{2003}), \eprint{0207047}.

\bibitem[{\citenamefont{Ralston and Pire}(2002)}]{Ralston:2001xs}
\bibinfo{author}{\bibfnamefont{J.~P.} \bibnamefont{Ralston}} \bibnamefont{and} \bibinfo{author}{\bibfnamefont{B.}~\bibnamefont{Pire}}, \bibinfo{journal}{Phys. Rev. D} \textbf{\bibinfo{volume}{66}}, \bibinfo{pages}{111501} (\bibinfo{year}{2002}), \eprint{hep-ph/0110075}.

\bibitem[{\citenamefont{Braun et~al.}(2012)\citenamefont{Braun, Manashov, and Pirnay}}]{Braun:2012bg}
\bibinfo{author}{\bibfnamefont{V.~M.} \bibnamefont{Braun}}, \bibinfo{author}{\bibfnamefont{A.~N.} \bibnamefont{Manashov}}, \bibnamefont{and} \bibinfo{author}{\bibfnamefont{B.}~\bibnamefont{Pirnay}}, \bibinfo{journal}{Phys. Rev. D} \textbf{\bibinfo{volume}{86}}, \bibinfo{pages}{014003} (\bibinfo{year}{2012}), \eprint{hep-ph/1205.3332}.

\bibitem[{\citenamefont{Braun et~al.}(2014)\citenamefont{Braun, Manashov, M\"uller, and Pirnay}}]{Braun:2014sta}
\bibinfo{author}{\bibfnamefont{V.~M.} \bibnamefont{Braun}}, \bibinfo{author}{\bibfnamefont{A.~N.} \bibnamefont{Manashov}}, \bibinfo{author}{\bibfnamefont{D.}~\bibnamefont{M\"uller}}, \bibnamefont{and} \bibinfo{author}{\bibfnamefont{B.~M.} \bibnamefont{Pirnay}}, \bibinfo{journal}{Phys. Rev. D} \textbf{\bibinfo{volume}{89}}, \bibinfo{pages}{074022} (\bibinfo{year}{2014}), \eprint{1401.7621}.

\bibitem[{\citenamefont{Braun et~al.}(2016)\citenamefont{Braun, Manashov, Moch, and Strohmaier}}]{Braun:2016qlg}
\bibinfo{author}{\bibfnamefont{V.~M.} \bibnamefont{Braun}}, \bibinfo{author}{\bibfnamefont{A.~N.} \bibnamefont{Manashov}}, \bibinfo{author}{\bibfnamefont{S.}~\bibnamefont{Moch}}, \bibnamefont{and} \bibinfo{author}{\bibfnamefont{M.}~\bibnamefont{Strohmaier}}, \bibinfo{journal}{JHEP} \textbf{\bibinfo{volume}{03}}, \bibinfo{pages}{142} (\bibinfo{year}{2016}), \eprint{1601.05937}.

\bibitem[{\citenamefont{Braun et~al.}(2021)\citenamefont{Braun, Ji, and Manashov}}]{Braun:2020zjm}
\bibinfo{author}{\bibfnamefont{V.~M.} \bibnamefont{Braun}}, \bibinfo{author}{\bibfnamefont{Y.}~\bibnamefont{Ji}}, \bibnamefont{and} \bibinfo{author}{\bibfnamefont{A.~N.} \bibnamefont{Manashov}}, \bibinfo{journal}{JHEP} \textbf{\bibinfo{volume}{03}}, \bibinfo{pages}{051} (\bibinfo{year}{2021}), \eprint{hep-ph/2011.04533}.

\bibitem[{\citenamefont{Braun et~al.}(2023)\citenamefont{Braun, Ji, and Manashov}}]{Braun:2022qly}
\bibinfo{author}{\bibfnamefont{V.~M.} \bibnamefont{Braun}}, \bibinfo{author}{\bibfnamefont{Y.}~\bibnamefont{Ji}}, \bibnamefont{and} \bibinfo{author}{\bibfnamefont{A.~N.} \bibnamefont{Manashov}}, \bibinfo{journal}{JHEP} \textbf{\bibinfo{volume}{01}}, \bibinfo{pages}{078} (\bibinfo{year}{2023}), \eprint{hep-ph/2211.04902}.

\bibitem[{\citenamefont{Radyushkin}(1999)}]{Radyushkin:1998bz}
\bibinfo{author}{\bibfnamefont{A.~V.} \bibnamefont{Radyushkin}}, \bibinfo{journal}{Phys. Lett. B} \textbf{\bibinfo{volume}{449}}, \bibinfo{pages}{81} (\bibinfo{year}{1999}), \eprint{hep-ph/9810466}.

\bibitem[{\citenamefont{Lorc\'e et~al.}(2022)\citenamefont{Lorc\'e, Pire, and Song}}]{Lorce:2022tiq}
\bibinfo{author}{\bibfnamefont{C.}~\bibnamefont{Lorc\'e}}, \bibinfo{author}{\bibfnamefont{B.}~\bibnamefont{Pire}}, \bibnamefont{and} \bibinfo{author}{\bibfnamefont{Q.-T.} \bibnamefont{Song}}, \bibinfo{journal}{Phys. Rev. D} \textbf{\bibinfo{volume}{106}}, \bibinfo{pages}{094030} (\bibinfo{year}{2022}), \eprint{2209.11140}.

\bibitem[{\citenamefont{Pire and Song}(2023)}]{Pire:2023kng}
\bibinfo{author}{\bibfnamefont{B.}~\bibnamefont{Pire}} \bibnamefont{and} \bibinfo{author}{\bibfnamefont{Q.-T.} \bibnamefont{Song}}, \bibinfo{journal}{Phys. Rev. D} \textbf{\bibinfo{volume}{107}}, \bibinfo{pages}{114014} (\bibinfo{year}{2023}), \eprint{2304.06389}.

\bibitem[{\citenamefont{Pire and Song}(2024)}]{Pire:2023ztb}
\bibinfo{author}{\bibfnamefont{B.}~\bibnamefont{Pire}} \bibnamefont{and} \bibinfo{author}{\bibfnamefont{Q.-T.} \bibnamefont{Song}}, \bibinfo{journal}{Phys. Rev. D} \textbf{\bibinfo{volume}{109}}, \bibinfo{pages}{074016} (\bibinfo{year}{2024}), \eprint{2311.06005}.

\bibitem[{\citenamefont{Schoenleber and Szafron}(2024)}]{Schoenleber:2024ihr}
\bibinfo{author}{\bibfnamefont{J.}~\bibnamefont{Schoenleber}} \bibnamefont{and} \bibinfo{author}{\bibfnamefont{R.}~\bibnamefont{Szafron}}, \bibinfo{journal}{JHEP} \textbf{\bibinfo{volume}{11}}, \bibinfo{pages}{031} (\bibinfo{year}{2024}), \eprint{2407.09263}.

\bibitem[{\citenamefont{Berger et~al.}(2002)\citenamefont{Berger, Diehl, and Pire}}]{Berger:2001xd}
\bibinfo{author}{\bibfnamefont{E.~R.} \bibnamefont{Berger}}, \bibinfo{author}{\bibfnamefont{M.}~\bibnamefont{Diehl}}, \bibnamefont{and} \bibinfo{author}{\bibfnamefont{B.}~\bibnamefont{Pire}}, \bibinfo{journal}{Eur. Phys. J.} \textbf{\bibinfo{volume}{C23}}, \bibinfo{pages}{675} (\bibinfo{year}{2002}), \eprint{hep-ph/0110062}.

\bibitem[{\citenamefont{Anikin et~al.}(2000)\citenamefont{Anikin, Pire, and Teryaev}}]{Anikin:2000em}
\bibinfo{author}{\bibfnamefont{I.~V.} \bibnamefont{Anikin}}, \bibinfo{author}{\bibfnamefont{B.}~\bibnamefont{Pire}}, \bibnamefont{and} \bibinfo{author}{\bibfnamefont{O.~V.} \bibnamefont{Teryaev}}, \bibinfo{journal}{Phys. Rev. D} \textbf{\bibinfo{volume}{62}}, \bibinfo{pages}{071501} (\bibinfo{year}{2000}), \eprint{hep-ph/0003203}.

\bibitem[{\citenamefont{Kivel et~al.}(2001)\citenamefont{Kivel, Polyakov, and Vanderhaeghen}}]{Kivel:2000fg}
\bibinfo{author}{\bibfnamefont{N.}~\bibnamefont{Kivel}}, \bibinfo{author}{\bibfnamefont{M.~V.} \bibnamefont{Polyakov}}, \bibnamefont{and} \bibinfo{author}{\bibfnamefont{M.}~\bibnamefont{Vanderhaeghen}}, \bibinfo{journal}{Phys. Rev. D} \textbf{\bibinfo{volume}{63}}, \bibinfo{pages}{114014} (\bibinfo{year}{2001}), \eprint{hep-ph/0012136}.

\bibitem[{\citenamefont{Radyushkin and Weiss}(2001)}]{Radyushkin:2000ap}
\bibinfo{author}{\bibfnamefont{A.~V.} \bibnamefont{Radyushkin}} \bibnamefont{and} \bibinfo{author}{\bibfnamefont{C.}~\bibnamefont{Weiss}}, \bibinfo{journal}{Phys. Rev. D} \textbf{\bibinfo{volume}{63}}, \bibinfo{pages}{114012} (\bibinfo{year}{2001}), \eprint{hep-ph/0010296}.

\bibitem[{\citenamefont{Aslan et~al.}(2018)\citenamefont{Aslan, Burkardt, Lorc\'e, Metz, and Pasquini}}]{Aslan:2018zzk}
\bibinfo{author}{\bibfnamefont{F.}~\bibnamefont{Aslan}}, \bibinfo{author}{\bibfnamefont{M.}~\bibnamefont{Burkardt}}, \bibinfo{author}{\bibfnamefont{C.}~\bibnamefont{Lorc\'e}}, \bibinfo{author}{\bibfnamefont{A.}~\bibnamefont{Metz}}, \bibnamefont{and} \bibinfo{author}{\bibfnamefont{B.}~\bibnamefont{Pasquini}}, \bibinfo{journal}{Phys. Rev. D} \textbf{\bibinfo{volume}{98}}, \bibinfo{pages}{014038} (\bibinfo{year}{2018}), \eprint{1802.06243}.

\bibitem[{\citenamefont{Belitsky and M{\"u}ller}(2003)}]{Belitsky:2003fj}
\bibinfo{author}{\bibfnamefont{A.~V.} \bibnamefont{Belitsky}} \bibnamefont{and} \bibinfo{author}{\bibfnamefont{D.}~\bibnamefont{M{\"u}ller}}, \bibinfo{journal}{Phys. Rev. D} \textbf{\bibinfo{volume}{68}}, \bibinfo{pages}{116005} (\bibinfo{year}{2003}), \eprint{hep-ph/0307369}.

\bibitem[{\citenamefont{Guidal and Vanderhaeghen}(2003)}]{guidal2003}
\bibinfo{author}{\bibfnamefont{M.}~\bibnamefont{Guidal}} \bibnamefont{and} \bibinfo{author}{\bibfnamefont{M.}~\bibnamefont{Vanderhaeghen}}, \bibinfo{journal}{Phys. Rev. Lett.} \textbf{\bibinfo{volume}{90}}, \bibinfo{pages}{012001} (\bibinfo{year}{2003}), \eprint{hep-ph/0208275}.

\bibitem[{\citenamefont{Deja et~al.}(2023)\citenamefont{Deja, Martinez-Fernandez, Pire, Sznajder, and Wagner}}]{Deja:2023ahc}
\bibinfo{author}{\bibfnamefont{K.}~\bibnamefont{Deja}}, \bibinfo{author}{\bibfnamefont{V.}~\bibnamefont{Martinez-Fernandez}}, \bibinfo{author}{\bibfnamefont{B.}~\bibnamefont{Pire}}, \bibinfo{author}{\bibfnamefont{P.}~\bibnamefont{Sznajder}}, \bibnamefont{and} \bibinfo{author}{\bibfnamefont{J.}~\bibnamefont{Wagner}}, \bibinfo{journal}{Phys. Rev. D} \textbf{\bibinfo{volume}{107}}, \bibinfo{pages}{094035} (\bibinfo{year}{2023}), \eprint{2303.13668}.

\bibitem[{\citenamefont{Alvarado et~al.}(2025)\citenamefont{Alvarado, Hoballah, and Voutier}}]{Alvarado:2025huq}
\bibinfo{author}{\bibfnamefont{J.~S.} \bibnamefont{Alvarado}}, \bibinfo{author}{\bibfnamefont{M.}~\bibnamefont{Hoballah}}, \bibnamefont{and} \bibinfo{author}{\bibfnamefont{E.}~\bibnamefont{Voutier}} (\bibinfo{year}{2025}), \eprint{2502.02346}.

\bibitem[{\citenamefont{Diehl}(2003)}]{diehl_review}
\bibinfo{author}{\bibfnamefont{M.}~\bibnamefont{Diehl}}, \bibinfo{journal}{Phys. Rept.} \textbf{\bibinfo{volume}{388}}, \bibinfo{pages}{41} (\bibinfo{year}{2003}), \eprint{hep-ph/0307382}.

\bibitem[{\citenamefont{Sullivan}(1972)}]{sullivanProcess}
\bibinfo{author}{\bibfnamefont{J.~D.} \bibnamefont{Sullivan}}, \bibinfo{journal}{Phys. Rev. D} \textbf{\bibinfo{volume}{5}}, \bibinfo{pages}{1732} (\bibinfo{year}{1972}).

\bibitem[{\citenamefont{Chatagnon et~al.}(2021)}]{CLAS:2021lky}
\bibinfo{author}{\bibfnamefont{P.}~\bibnamefont{Chatagnon}} \bibnamefont{et~al.}, \bibinfo{journal}{Phys. Rev. Lett.} \textbf{\bibinfo{volume}{127}}, \bibinfo{pages}{262501} (\bibinfo{year}{2021}), \eprint{hep-ex/2108.11746}.

\bibitem[{\citenamefont{Grocholski et~al.}(2020)\citenamefont{Grocholski, Moutarde, Pire, Sznajder, and Wagner}}]{Grocholski:2019pqj}
\bibinfo{author}{\bibfnamefont{O.}~\bibnamefont{Grocholski}}, \bibinfo{author}{\bibfnamefont{H.}~\bibnamefont{Moutarde}}, \bibinfo{author}{\bibfnamefont{B.}~\bibnamefont{Pire}}, \bibinfo{author}{\bibfnamefont{P.}~\bibnamefont{Sznajder}}, \bibnamefont{and} \bibinfo{author}{\bibfnamefont{J.}~\bibnamefont{Wagner}}, \bibinfo{journal}{Eur. Phys. J. C} \textbf{\bibinfo{volume}{80}}, \bibinfo{pages}{171} (\bibinfo{year}{2020}), \eprint{1912.09853}.

\bibitem[{\citenamefont{Mueller et~al.}(2012)\citenamefont{Mueller, Pire, Szymanowski, and Wagner}}]{Mueller:2012sma}
\bibinfo{author}{\bibfnamefont{D.}~\bibnamefont{Mueller}}, \bibinfo{author}{\bibfnamefont{B.}~\bibnamefont{Pire}}, \bibinfo{author}{\bibfnamefont{L.}~\bibnamefont{Szymanowski}}, \bibnamefont{and} \bibinfo{author}{\bibfnamefont{J.}~\bibnamefont{Wagner}}, \bibinfo{journal}{Phys. Rev. D} \textbf{\bibinfo{volume}{86}}, \bibinfo{pages}{031502} (\bibinfo{year}{2012}), \eprint{1203.4392}.

\bibitem[{\citenamefont{Chen et~al.}(2014)\citenamefont{Chen, Gao, Hemmick, Meziani, and Souder}}]{Chen:2014psa}
\bibinfo{author}{\bibfnamefont{J.~P.} \bibnamefont{Chen}}, \bibinfo{author}{\bibfnamefont{H.}~\bibnamefont{Gao}}, \bibinfo{author}{\bibfnamefont{T.~K.} \bibnamefont{Hemmick}}, \bibinfo{author}{\bibfnamefont{Z.~E.} \bibnamefont{Meziani}}, \bibnamefont{and} \bibinfo{author}{\bibfnamefont{P.~A.} \bibnamefont{Souder}} (\bibinfo{year}{2014}), \bibinfo{note}{{A White Paper on SoLID (Solenoidal Large Intensity Device)}}, \eprint{nucl-ex/1409.7741}.

\bibitem[{\citenamefont{Camsonne}(2017)}]{Camsonne:2017yux}
\bibinfo{author}{\bibfnamefont{A.}~\bibnamefont{Camsonne}}, \bibinfo{journal}{PoS} \textbf{\bibinfo{volume}{INPC2016}}, \bibinfo{pages}{309} (\bibinfo{year}{2017}).

\bibitem[{\citenamefont{Zhao et~al.}(2021)\citenamefont{Zhao, Camsonne, Marchand, Mazouz, Sparveris, Stepanyan, Voutier, and Zhao}}]{Zhao:2021zsm}
\bibinfo{author}{\bibfnamefont{S.}~\bibnamefont{Zhao}}, \bibinfo{author}{\bibfnamefont{A.}~\bibnamefont{Camsonne}}, \bibinfo{author}{\bibfnamefont{D.}~\bibnamefont{Marchand}}, \bibinfo{author}{\bibfnamefont{M.}~\bibnamefont{Mazouz}}, \bibinfo{author}{\bibfnamefont{N.}~\bibnamefont{Sparveris}}, \bibinfo{author}{\bibfnamefont{S.}~\bibnamefont{Stepanyan}}, \bibinfo{author}{\bibfnamefont{E.}~\bibnamefont{Voutier}}, \bibnamefont{and} \bibinfo{author}{\bibfnamefont{Z.~W.} \bibnamefont{Zhao}}, \bibinfo{journal}{Eur. Phys. J. A} \textbf{\bibinfo{volume}{57}}, \bibinfo{pages}{240} (\bibinfo{year}{2021}), \eprint{nucl-ex/2103.12773}.

\bibitem[{\citenamefont{Accardi et~al.}(2024)}]{Accardi:2023chb}
\bibinfo{author}{\bibfnamefont{A.}~\bibnamefont{Accardi}} \bibnamefont{et~al.}, \bibinfo{journal}{Eur. Phys. J. A} \textbf{\bibinfo{volume}{60}}, \bibinfo{pages}{173} (\bibinfo{year}{2024}), \eprint{2306.09360}.

\bibitem[{\citenamefont{Abdul~Khalek et~al.}(2022)}]{AbdulKhalek:2021gbh}
\bibinfo{author}{\bibfnamefont{R.}~\bibnamefont{Abdul~Khalek}} \bibnamefont{et~al.}, \bibinfo{journal}{Nucl. Phys. A} \textbf{\bibinfo{volume}{1026}}, \bibinfo{pages}{122447} (\bibinfo{year}{2022}), \eprint{physics.ins-det/2103.05419}.

\bibitem[{\citenamefont{Anderle et~al.}(2021)}]{Anderle:2021wcy}
\bibinfo{author}{\bibfnamefont{D.~P.} \bibnamefont{Anderle}} \bibnamefont{et~al.}, \bibinfo{journal}{Front. Phys. (Beijing)} \textbf{\bibinfo{volume}{16}}, \bibinfo{pages}{64701} (\bibinfo{year}{2021}), \eprint{nucl-ex/2102.09222}.

\bibitem[{\citenamefont{Chavez et~al.}(2022)\citenamefont{Chavez, Bertone, De~Soto~Borrero, Defurne, Mezrag, Moutarde, Rodr\'\i{}guez-Quintero, and Segovia}}]{Chavez:2021llq}
\bibinfo{author}{\bibfnamefont{J.~M.~M.} \bibnamefont{Chavez}}, \bibinfo{author}{\bibfnamefont{V.}~\bibnamefont{Bertone}}, \bibinfo{author}{\bibfnamefont{F.}~\bibnamefont{De~Soto~Borrero}}, \bibinfo{author}{\bibfnamefont{M.}~\bibnamefont{Defurne}}, \bibinfo{author}{\bibfnamefont{C.}~\bibnamefont{Mezrag}}, \bibinfo{author}{\bibfnamefont{H.}~\bibnamefont{Moutarde}}, \bibinfo{author}{\bibfnamefont{J.}~\bibnamefont{Rodr\'\i{}guez-Quintero}}, \bibnamefont{and} \bibinfo{author}{\bibfnamefont{J.}~\bibnamefont{Segovia}}, \bibinfo{journal}{Phys. Rev. D} \textbf{\bibinfo{volume}{105}}, \bibinfo{pages}{094012} (\bibinfo{year}{2022}), \eprint{2110.06052}.

\bibitem[{\citenamefont{Martinez-Fernandez et~al.}(2025)\citenamefont{Martinez-Fernandez, Pire, Sznajder, and Wagner}}]{Martinez-Fernandez:2024pzm}
\bibinfo{author}{\bibfnamefont{V.}~\bibnamefont{Martinez-Fernandez}}, \bibinfo{author}{\bibfnamefont{B.}~\bibnamefont{Pire}}, \bibinfo{author}{\bibfnamefont{P.}~\bibnamefont{Sznajder}}, \bibnamefont{and} \bibinfo{author}{\bibfnamefont{J.}~\bibnamefont{Wagner}}, \bibinfo{journal}{PoS} \textbf{\bibinfo{volume}{DIS2024}}, \bibinfo{pages}{205} (\bibinfo{year}{2025}), \eprint{2406.14640}.

\bibitem[{\citenamefont{Martinez-Fernández}(2024)}]{vthesis}
\bibinfo{author}{\bibfnamefont{V.}~\bibnamefont{Martinez-Fernández}}, \bibinfo{type}{{Ph.D. thesis}}, \bibinfo{school}{National Centre for Nuclear Research, Poland} (\bibinfo{year}{2024}).

\bibitem[{\citenamefont{Di~Francesco et~al.}(1997)\citenamefont{Di~Francesco, Mathieu, and Senechal}}]{difrancesco}
\bibinfo{author}{\bibfnamefont{P.}~\bibnamefont{Di~Francesco}}, \bibinfo{author}{\bibfnamefont{P.}~\bibnamefont{Mathieu}}, \bibnamefont{and} \bibinfo{author}{\bibfnamefont{D.}~\bibnamefont{Senechal}}, \emph{\bibinfo{title}{{Conformal Field Theory}}}, Graduate Texts in Contemporary Physics (\bibinfo{publisher}{Springer-Verlag}, \bibinfo{address}{New York}, \bibinfo{year}{1997}), ISBN \bibinfo{isbn}{978-0-387-94785-3, 978-1-4612-7475-9}.

\bibitem[{\citenamefont{Ferrara et~al.}(1972{\natexlab{a}})\citenamefont{Ferrara, Grillo, Parisi, and Gatto}}]{ferrara1972}
\bibinfo{author}{\bibfnamefont{S.}~\bibnamefont{Ferrara}}, \bibinfo{author}{\bibfnamefont{A.~F.} \bibnamefont{Grillo}}, \bibinfo{author}{\bibfnamefont{G.}~\bibnamefont{Parisi}}, \bibnamefont{and} \bibinfo{author}{\bibfnamefont{R.}~\bibnamefont{Gatto}}, \bibinfo{journal}{Lett. Nuovo Cim.} \textbf{\bibinfo{volume}{4S2}}, \bibinfo{pages}{115} (\bibinfo{year}{1972}{\natexlab{a}}).

\bibitem[{\citenamefont{Ferrara and Parisi}(1972)}]{ferraraparisi1972}
\bibinfo{author}{\bibfnamefont{S.}~\bibnamefont{Ferrara}} \bibnamefont{and} \bibinfo{author}{\bibfnamefont{G.}~\bibnamefont{Parisi}}, \bibinfo{journal}{Nuclear physics B} \textbf{\bibinfo{volume}{42}}, \bibinfo{pages}{281} (\bibinfo{year}{1972}).

\bibitem[{\citenamefont{Ferrara et~al.}(1972{\natexlab{b}})\citenamefont{Ferrara, Grillo, Parisi, and Gatto}}]{ferraragrilloparisigatto1972}
\bibinfo{author}{\bibfnamefont{S.}~\bibnamefont{Ferrara}}, \bibinfo{author}{\bibfnamefont{A.~F.} \bibnamefont{Grillo}}, \bibinfo{author}{\bibfnamefont{G.}~\bibnamefont{Parisi}}, \bibnamefont{and} \bibinfo{author}{\bibfnamefont{R.}~\bibnamefont{Gatto}}, \bibinfo{journal}{Physics Letters B} \textbf{\bibinfo{volume}{38}}, \bibinfo{pages}{333} (\bibinfo{year}{1972}{\natexlab{b}}).

\bibitem[{\citenamefont{Geyer et~al.}(1999)\citenamefont{Geyer, Lazar, and Robaschik}}]{geyer1999}
\bibinfo{author}{\bibfnamefont{B.}~\bibnamefont{Geyer}}, \bibinfo{author}{\bibfnamefont{M.}~\bibnamefont{Lazar}}, \bibnamefont{and} \bibinfo{author}{\bibfnamefont{D.}~\bibnamefont{Robaschik}}, \bibinfo{journal}{Nucl. Phys. B} \textbf{\bibinfo{volume}{559}}, \bibinfo{pages}{339} (\bibinfo{year}{1999}), \eprint{hep-th/9901090}.

\bibitem[{\citenamefont{Belitsky et~al.}(2014)\citenamefont{Belitsky, M\"uller, and Ji}}]{Belitsky:2012ch}
\bibinfo{author}{\bibfnamefont{A.~V.} \bibnamefont{Belitsky}}, \bibinfo{author}{\bibfnamefont{D.}~\bibnamefont{M\"uller}}, \bibnamefont{and} \bibinfo{author}{\bibfnamefont{Y.}~\bibnamefont{Ji}}, \bibinfo{journal}{Nucl. Phys. B} \textbf{\bibinfo{volume}{878}}, \bibinfo{pages}{214} (\bibinfo{year}{2014}), \eprint{1212.6674}.

\bibitem[{\citenamefont{Srednicki}(2007)}]{srednicki}
\bibinfo{author}{\bibfnamefont{M.}~\bibnamefont{Srednicki}}, \emph{\bibinfo{title}{{Quantum field theory}}} (\bibinfo{publisher}{Cambridge University Press}, \bibinfo{year}{2007}), ISBN \bibinfo{isbn}{978-0-521-86449-7, 978-0-511-26720-8}.

\bibitem[{\citenamefont{Tarrach}(1975)}]{Tarrach:1975tu}
\bibinfo{author}{\bibfnamefont{R.}~\bibnamefont{Tarrach}}, \bibinfo{journal}{Nuovo Cim. A} \textbf{\bibinfo{volume}{28}}, \bibinfo{pages}{409} (\bibinfo{year}{1975}).

\bibitem[{\citenamefont{Belitsky and M{\"u}ller}(2000)}]{Belitsky:2000jk}
\bibinfo{author}{\bibfnamefont{A.~V.} \bibnamefont{Belitsky}} \bibnamefont{and} \bibinfo{author}{\bibfnamefont{D.}~\bibnamefont{M{\"u}ller}}, \bibinfo{journal}{Phys. Lett. B} \textbf{\bibinfo{volume}{486}}, \bibinfo{pages}{369} (\bibinfo{year}{2000}), \eprint{hep-ph/0005028}.

\bibitem[{\citenamefont{Musatov and Radyushkin}(2000)}]{Musatov:1999xp}
\bibinfo{author}{\bibfnamefont{I.~V.} \bibnamefont{Musatov}} \bibnamefont{and} \bibinfo{author}{\bibfnamefont{A.~V.} \bibnamefont{Radyushkin}}, \bibinfo{journal}{Phys. Rev. D} \textbf{\bibinfo{volume}{61}}, \bibinfo{pages}{074027} (\bibinfo{year}{2000}), \eprint{hep-ph/9905376}.

\bibitem[{\citenamefont{Novikov et~al.}(2020)}]{Novikov:2020snp}
\bibinfo{author}{\bibfnamefont{I.}~\bibnamefont{Novikov}} \bibnamefont{et~al.}, \bibinfo{journal}{Phys. Rev. D} \textbf{\bibinfo{volume}{102}}, \bibinfo{pages}{014040} (\bibinfo{year}{2020}), \eprint{2002.02902}.

\bibitem[{\citenamefont{Amrath et~al.}(2008)\citenamefont{Amrath, Diehl, and Lansberg}}]{Amrath:2008vx}
\bibinfo{author}{\bibfnamefont{D.}~\bibnamefont{Amrath}}, \bibinfo{author}{\bibfnamefont{M.}~\bibnamefont{Diehl}}, \bibnamefont{and} \bibinfo{author}{\bibfnamefont{J.-P.} \bibnamefont{Lansberg}}, \bibinfo{journal}{Eur. Phys. J. C} \textbf{\bibinfo{volume}{58}}, \bibinfo{pages}{179} (\bibinfo{year}{2008}), \eprint{0807.4474}.

\bibitem[{\citenamefont{Berger et~al.}(2001)\citenamefont{Berger, Cano, Diehl, and Pire}}]{Berger:2001zb}
\bibinfo{author}{\bibfnamefont{E.~R.} \bibnamefont{Berger}}, \bibinfo{author}{\bibfnamefont{F.}~\bibnamefont{Cano}}, \bibinfo{author}{\bibfnamefont{M.}~\bibnamefont{Diehl}}, \bibnamefont{and} \bibinfo{author}{\bibfnamefont{B.}~\bibnamefont{Pire}}, \bibinfo{journal}{Phys. Rev. Lett.} \textbf{\bibinfo{volume}{87}}, \bibinfo{pages}{142302} (\bibinfo{year}{2001}), \eprint{hep-ph/0106192}.

\bibitem[{\citenamefont{Cano and Pire}(2004)}]{Cano:2003ju}
\bibinfo{author}{\bibfnamefont{F.}~\bibnamefont{Cano}} \bibnamefont{and} \bibinfo{author}{\bibfnamefont{B.}~\bibnamefont{Pire}}, \bibinfo{journal}{Eur. Phys. J. A} \textbf{\bibinfo{volume}{19}}, \bibinfo{pages}{423} (\bibinfo{year}{2004}), \eprint{hep-ph/0307231}.

\bibitem[{\citenamefont{Hobart et~al.}(2024)}]{CLAS:2024qhy}
\bibinfo{author}{\bibfnamefont{A.}~\bibnamefont{Hobart}} \bibnamefont{et~al.} (\bibinfo{collaboration}{CLAS}), \bibinfo{journal}{Phys. Rev. Lett.} \textbf{\bibinfo{volume}{133}}, \bibinfo{pages}{211903} (\bibinfo{year}{2024}), \eprint{2406.15539}.

\bibitem[{\citenamefont{Moutarde et~al.}(2013)\citenamefont{Moutarde, Pire, Sabatie, Szymanowski, and Wagner}}]{Moutarde:2013qs}
\bibinfo{author}{\bibfnamefont{H.}~\bibnamefont{Moutarde}}, \bibinfo{author}{\bibfnamefont{B.}~\bibnamefont{Pire}}, \bibinfo{author}{\bibfnamefont{F.}~\bibnamefont{Sabatie}}, \bibinfo{author}{\bibfnamefont{L.}~\bibnamefont{Szymanowski}}, \bibnamefont{and} \bibinfo{author}{\bibfnamefont{J.}~\bibnamefont{Wagner}}, \bibinfo{journal}{Phys. Rev. D} \textbf{\bibinfo{volume}{87}}, \bibinfo{pages}{054029} (\bibinfo{year}{2013}), \eprint{1301.3819}.

\bibitem[{\citenamefont{Bhattacharya et~al.}(2023)\citenamefont{Bhattacharya, Cichy, Constantinou, Dodson, Metz, Scapellato, and Steffens}}]{Bhattacharya:2023nmv}
\bibinfo{author}{\bibfnamefont{S.}~\bibnamefont{Bhattacharya}}, \bibinfo{author}{\bibfnamefont{K.}~\bibnamefont{Cichy}}, \bibinfo{author}{\bibfnamefont{M.}~\bibnamefont{Constantinou}}, \bibinfo{author}{\bibfnamefont{J.}~\bibnamefont{Dodson}}, \bibinfo{author}{\bibfnamefont{A.}~\bibnamefont{Metz}}, \bibinfo{author}{\bibfnamefont{A.}~\bibnamefont{Scapellato}}, \bibnamefont{and} \bibinfo{author}{\bibfnamefont{F.}~\bibnamefont{Steffens}}, \bibinfo{journal}{Phys. Rev. D} \textbf{\bibinfo{volume}{108}}, \bibinfo{pages}{054501} (\bibinfo{year}{2023}), \eprint{2306.05533}.

\bibitem[{\citenamefont{Dupr\'e et~al.}(2021)}]{CLAS:2021ovm}
\bibinfo{author}{\bibfnamefont{R.}~\bibnamefont{Dupr\'e}} \bibnamefont{et~al.} (\bibinfo{collaboration}{CLAS}), \bibinfo{journal}{Phys. Rev. C} \textbf{\bibinfo{volume}{104}}, \bibinfo{pages}{025203} (\bibinfo{year}{2021}), \eprint{2102.07419}.

\bibitem[{\citenamefont{Polyakov and Weiss}(1999)}]{Polyakov:1999gs}
\bibinfo{author}{\bibfnamefont{M.~V.} \bibnamefont{Polyakov}} \bibnamefont{and} \bibinfo{author}{\bibfnamefont{C.}~\bibnamefont{Weiss}}, \bibinfo{journal}{Phys. Rev. D} \textbf{\bibinfo{volume}{60}}, \bibinfo{pages}{114017} (\bibinfo{year}{1999}), \eprint{hep-ph/9902451}.

\end{thebibliography}
